\documentclass[aip,reprint]{revtex4-1}

\usepackage{graphicx}  
\usepackage{dcolumn}   
\usepackage{bm}        
\usepackage{amssymb}   
\usepackage{amsmath}
\usepackage{comment}
\usepackage[utf8]{inputenc}
\usepackage{float}
\usepackage[dvipsnames]{xcolor}
\usepackage{graphicx}
\usepackage{comment}
\usepackage{bigints}
\newcommand{\usigma}{\bm{\sigma}}
\newcommand{\pltime}{\tau_{\rm{pl}}}
\newcommand{\ps}{P^{\rm ss}}
\newcommand{\E}{q_0}
\newcommand{\etau}{\delta \sigma_{\rm{u}}}
\newcommand{\mfpt}{\tau_{\rm{FP}}}
\newcommand{\betaf}{\mathrm{B}}
\newcommand{\So}{S_0}
\newcommand{\Rext}{R_1^{\rm{ext}}}
\newcommand{\Rint}{R_1^{\rm{int}}}
\newcommand{\prefaging}{B}
\newcommand{\prefexp}{C_1}

\newcommand{\Go}{G_0}
\newcommand{\dplas}{r_0}
\newcommand{\aplas}{a}
\newcommand{\R}{R}
\newcommand{\pack}{\varphi}
\newcommand{\abso}{\phi}
\newcommand{\prefQ}{B}

\draft 

\begin{document}


\title{Aging in a mean field elastoplastic model of amorphous solids} 



\author{Jack T. Parley}
 \email[Author to whom correspondence should be addressed: ]{jack.parley@uni-goettingen.de}
 \affiliation{Institut f{\"u}r Theoretische Physik, University of G{\"o}ttingen,
Friedrich-Hund-Platz 1, 37077 G{\"o}ttingen, Germany}


\author{Suzanne M. Fielding}%
\affiliation{ 
Department of Physics, Durham University, Science Laboratories, South Road, Durham DH1 3LE, UK}

\author{Peter Sollich}
\affiliation{Institut f{\"u}r Theoretische Physik, University of G{\"o}ttingen,
Friedrich-Hund-Platz 1, 37077 G{\"o}ttingen, Germany}
\affiliation{Department of Mathematics, King’s College London, London WC2R 2LS, UK}



\date{\today}

\begin{abstract}
We construct a mean-field elastoplastic description of the dynamics of amorphous solids under arbitrary time-dependent perturbations, building on the work of Lin and Wyart [J. Lin and M. Wyart, Phys. Rev. X 6, 011005 (2016)] for steady shear. Local stresses are driven by power-law distributed mechanical noise from yield events throughout the material,
in contrast to the well-studied H\'{e}braud-Lequeux model where the noise is Gaussian.
We first use a mapping to a mean first passage time problem to study the phase diagram in the absence of shear, which shows a transition between an arrested and a fluid state. We then introduce a boundary layer scaling technique for low yield rate regimes, which we first apply to study the scaling of the steady state yield rate on approaching the arrest transition. These scalings are further developed to study the {\em aging} behaviour in the glassy regime, for different values of the exponent $\mu$ characterizing the mechanical noise spectrum. We find that the yield rate decays as a power-law for $1<\mu<2$, a stretched exponential for $\mu=1$ and an exponential for $\mu<1$, reflecting the relative importance of far-field and near-field events as the range of the stress propagator is varied. Comparison of the mean-field predictions with aging simulations of a lattice elastoplastic model shows excellent quantitative agreement, up to a simple rescaling of time.
\end{abstract}


\maketitle 

\section{\label{sec:level1}Introduction}

Amorphous materials, ranging from granular materials to foams and emulsions, show rich and complex behaviour under deformation and flow~\cite{nicolas_deformation_2018,bonn_yield_2017}. An important step both for fundamental understanding and for practical applications has been the development of elastoplastic models, which consider mesoscopic blocks of material, large enough so that an elastic shear stress can be ascribed to them. These local stresses have a dynamics that alternates between loading by external shear and plastic relaxation. Each plastic stress relaxation also affects stresses in the rest of the material through the Eshelby stress propagator, potentially triggering other plastic events. In a mean-field setting this effect can be modelled as \textit{mechanical noise}; in the athermal regime the strength of this noise is coupled directly to the level of plastic activity in the system. In pioneering work by H\'{e}braud and Lequeux~\cite{hebraud_mode-coupling_1998} (HL) the mechanical noise was modelled as Gaussian white noise, leading to a diffusive dynamics of the local stresses. Despite its simplicity, the model manages to predict a transition between a fluid and an arrested state, reproducing in the latter regime the Herschel-Bulkley law, which fits well the stationary flow behaviour of many yield stress materials.

To go beyond the HL approximation one needs to take into account the actual spatial decay of the stress field arising from an isolated plastic event~\cite{picard_elastic_2004}. Assuming that events occur randomly in space gives then a mechanical noise distribution that is not Gaussian but instead follows a power law~\cite{lemaitre_plastic_2007-1,lemaitre_plastic_2007}. This in turn leads to anomalous diffusive dynamics in stress. In the work of Lin and Wyart~\cite{lin_mean-field_2016}, a mean-field model is developed along these lines. The results from this study for the exponents associated with the yielding transition suggest this is the ``true'' mean-field model in the sense that it applies in large dimensions~\cite{lin_mean-field_2016}. However, so far the model has only been studied in quasistatic shear~\cite{lin_mean-field_2016,lin_microscopic_2018}, and an extension to general time-dependent aging and rheological phenomena is lacking. Having this time-dependent model will be important to test the mean-field assumptions against a full range of rheological experiments. The aim of the present work is to construct such an extension and to explore its dynamical properties, focussing on scenarios without external shear perturbation. These are relevant for studying to what extent athermal aging (such as in Ref.~\onlinecite{chacko_slow_2019}) can be described in a mean-field setting, and a first step towards comparing with rheological experiments where a perturbation is applied only after a given waiting time~\cite{agarwal_signatures_2019,lidon_power-law_2017,purnomo_glass_2008,purnomo_linear_2006,purnomo_rheological_2007}.

The paper is structured as follows. In Section~\ref{sec:model_construction} we detail the construction of the fully time-dependent mean-field elastoplastic model. We then, in Section~\ref{sec:phase_diagram}, determine the phase diagram of the model, separating the arrested (i.e.\ glassy) and flowing (liquid) states. In the following Section~\ref{sec:boundary_layer} we introduce a boundary layer scaling technique for the regime of low yield rates, building on the approach for the HL model~\cite{sollich_aging_2017}. The method is first applied to find the scaling of the yield rate above the transition, in Section~\ref{sec:scaling}. Then we deploy the boundary layer scaling technique to study the aging behaviour, in Section~\ref{sec:aging}. Finally, we compare the mean-field predictions and asymptotic behaviour to simulations of a lattice elastoplastic model in Section~\ref{sec:lattice}, and discuss the results and outlook towards future research in the final Section~\ref{sec:outlook}.

\section{\label{sec:model_construction}Time-dependent mean-field elastoplastic model}

 Following the general philosophy of elastoplastic models, we regard our system of interest as divided into $N$ mesoscopic blocks centred on the sites of a regular (e.g.\ square) lattice; to each block we assign a local shear stress. We begin with a description  of the stochastic rules governing the local stress dynamics. Considering  initially dynamics in discrete time, we introduce the following update rules. Take the time step $\Delta t$ small enough so that there is at most one yield event per time interval, and label the site where this event takes place by $l$. The yielding rule is given by
\begin{equation}
\sigma_l(t+\Delta t)=0 \quad \text{with probability} \  \frac{\Delta t }{\pltime} \quad \text{if}\  |\sigma_l(t)|>\sigma_c 
\label{sigma_l_update}
\end{equation}
This means that a yield event, where particles rearrange plastically, resets the local stress to zero~\footnote{We consider here a full local stress relaxation. The post-yield local stress may however also be modelled as drawn from a distribution of residual stress other than a Dirac function at zero. Previous studies have shown this not to change the behaviour qualitatively~\cite{agoritsas_non-trivial_2017}.}. Such a plastic rearrangement takes place at a fixed rate $\pltime^{-1}$ once the local stress exceeds the local yield threshold $\sigma_c$.

The stresses at all other sites $\{\sigma_i\}$, $i\neq l$, evolve as
\begin{equation}
  	\sigma_i(t+\Delta t)=G_0\dot{\gamma}\Delta t+\sigma_i(t)+\delta \sigma_i
  	\label{sigma_i_update}
\end{equation}
This incorporates a drift term due to the external shear rate $\dot{\gamma}$, multiplied by the local shear modulus $G_0$, and a stress ``kick'' $\delta \sigma_i$ that models the Eshelby stress propagation from the yield event at site $l$; $\delta\sigma_i$ of course depends on $l$ but we do not write this explicitly.
From the beginning we consider here a mean-field description of this mechanical noise, which can be derived from the full spatial Eshelby stress propagator~\cite{eshelby_determination_1957,picard_elastic_2004} by treating the site of the yield event as randomly chosen across the system. It is then straightforward to show~\cite{lin_mean-field_2016} that for a stress propagator decaying as $\sim r^{-\beta}$ in dimension $d$, the noise kicks will be distributed as $\rho(\delta \sigma)\sim |\delta \sigma|^{-\mu-1}$ with an exponent
\begin{equation}
    \mu=\frac{d}{\beta}
\end{equation}
For the physical propagator with $\beta=d$ (Ref.~\onlinecite{picard_elastic_2004}) this leads to $\mu=1$. We note though that recent works~\cite{fernandez_aguirre_critical_2018,ferrero_criticality_2019} suggest that exponent values in the range $1<\mu<2$ may also have physical relevance, once the description is coarse-grained further to study the aggregate effect of avalanches of yield events that can potentially span a large number of sites. Studying this range is also important in itself due to the marginal character of $\mu=1$, which will need to be approached as a limiting case. 
Varying $\mu$ can be thought of as varying the range of the stress propagation: larger $\mu$ corresponds to smaller $\beta$ and hence longer range propagation. In fact for $\mu\to 2$ we will find that our model reduces to the HL model with its effectively infinite interaction range. Conversely, for $\mu<1$ stress propagation is essentially local.

Besides the exponent $\mu$, the second key parameter of the model is the coupling $A$, related to the prefactor of the aforementioned power-law behaviour of $\rho(\delta \sigma)$. We can distinguish here two different approaches (see also the discussion, Sec.~\ref{sec:outlook}). Firstly, shown in App.~\ref{app:circular_geometry} is a derivation of the coupling in a $2$D setting of randomly distributed sites, leading to the result that the coupling $A$ depends both on the strength of the elastic interactions and on the density of sites where events may take place (see Eq.~\ref{rho_from_circle}). In a second approach (Sec.~\ref{sec:lattice}), where we consider instead the case of sites fixed to positions on a lattice, we will derive the value of $A$ by fitting directly the histogram of stress increments $\{\delta \sigma\}$, this value of $A$ being fixed by geometry. To include both cases, we treat $A$ as a variable parameter in the following.

The power law behaviour  $\rho(\delta \sigma)\sim |\delta \sigma|^{-\mu-1}$ derived above  is exact for small $\delta\sigma$, corresponding to the small effects of far away yield events, but must eventually be cut off at the largest  $|\delta\sigma|$ resulting from yield events at neighbouring sites. The explicit calculation in $d=2$
(Appendix~\ref{app:circular_geometry}) that accounts for the full angular dependence of the stress propagator gives a soft upper cutoff, where $\rho(\delta\sigma)$ goes to zero continuously.

For simplicity~\footnote{This will change quantitative aspects such as the location of the phase diagram; however, the scaling forms below are determined by the asymptotic power-law regime for small $|\delta\sigma|$, making the precise form of the upper cutoff irrelevant.} we nonetheless use for the calculations a hard upper cutoff $\etau$ as proposed in Ref.~\onlinecite{lin_mean-field_2016}, and a lower cutoff chosen as $\etau N^{-1/\mu}$ that goes to zero for $N\gg1$. This resulting simplified mechanical noise distribution takes the form
\begin{eqnarray}
\rho (\delta \sigma) = \frac{A}{N}|\delta \sigma |^{-\mu-1}, \quad  \delta \sigma_{u}N^{-1/\mu}<|\delta\sigma|<\etau \label{rho_wyart} \end{eqnarray}
By normalization, $\etau$ is then related to the prefactor (coupling) $A$ by
\begin{eqnarray}\label{upper_cutoff}
 \etau &= & \left(\frac{2A}{\mu}\right)^{1/\mu}\label{etau_wyart}
\end{eqnarray}
It is important to note that the form (\ref{upper_cutoff}) relies on a specific choice of the ratio between upper and lower cutoff. This was taken as $N^{1/\mu}$ in Ref.~\onlinecite{lin_mean-field_2016}, but in general involves a geometrical factor that will depend on the system. This means that (even within the assumption of a hard upper cutoff), the value of $\etau$ is not unambiguously fixed by $A$, a fact that we will return to in Sec.~\ref{sec:lattice}.

From the noise distribution (\ref{rho_wyart}) we extract, after a yield event at site $l$, independently and identically distributed (i.i.d.) stress increments $\delta\sigma_i$ for all other sites ($i\neq l$). We add to each $\delta\sigma_i$ in (\ref{sigma_i_update}) the term $-\sum_{k \neq l}\delta \sigma_k/(N-1)$. This counterterm is formally necessary to ensure that stress propagation has a net zero effect on the total stress $\sum_{i\neq l}\sigma_i$ outside the block that yields. For large $N$ we will see that the counterterm has a negligible effect because due to the symmetry of $\rho(\delta\sigma)$, $\sum_{k\neq l}\delta\sigma_k/(N-1)$ is of order $N^{1/2}/N=N^{-1/2}$.

We next transform the above dynamical rules into a master equation for the joint time evolution of the stresses $\bm{\sigma}=(\sigma_1,\ldots,\sigma_N)$ at all $N$ sites.
Given the above assumptions, 
the transition rate from configuration $\usigma'$ to $\usigma$ associated to a yield event at site $l$ is
\begin{multline}\label{rate}
    K_l(\usigma|\usigma')=\frac{1}{\pltime}
    \theta(|\sigma_l'|-\sigma_c)\delta(\sigma_l) \\ \left\langle \prod \limits_{j \neq l}\delta\left(\sigma_j-\left(\sigma_j'+\delta \sigma_j-\frac{1}{N-1}\sum \limits_{k \neq l}\delta \sigma_k\right)\right)\right\rangle
\end{multline}
where the factors in the first and second line correspond to (\ref{sigma_l_update}) and (\ref{sigma_i_update}), respectively. The brackets denote an average over the distribution of stress kicks $\delta\sigma_i$, which we recall are sampled independently from the same distribution $\rho(\delta \sigma)$.
Bearing in mind that a yield event can occur at any site $l$ and incorporating the loading by external shear in (\ref{sigma_i_update}) gives then the master equation 
\begin{multline}\label{N_body}
  \partial_t P(\usigma)=-G_0\dot{\gamma}\sum \limits_{i}\partial_{\sigma_i}P(\usigma)\\+\sum \limits_{l}\int \left(K_l(\usigma|\usigma')P(\usigma')-K_l(\usigma'|\usigma)P(\usigma)\right)\mathrm{d}\usigma'
\end{multline}

We can now reduce this description to one for the distribution of {\em local} stresses,
proceeding in a similar fashion to Ref.~\onlinecite{bocquet_kinetic_2009}. We assume a mean-field factorization $P(\usigma)=\prod \limits_{i}P_i(\sigma_i)$, which we expect to become exact for $N \rightarrow \infty$ as each local stress couples to the others only via the total number of yield events. From equation (\ref{N_body}) one can then obtain the time evolution of $P_i(\sigma_i)$ by integrating out the remaining $N-1$ stresses (see Appendix~\ref{app:model_construction}), which will include the effect of stress kicks from yield events at other sites. The final form of the master equation for the local stress distribution $P(\sigma)=(1/N)\sum_i\langle \delta(\sigma-\sigma_i)\rangle  = (1/N)\sum_i P_i(\sigma)$ reads
 \begin{eqnarray}\label{pde}
     \partial_t P(\sigma,t)&=&-G_0\dot{\gamma}\partial_{\sigma}P(\sigma,t)
     \nonumber\\
     &&{}+A \Gamma(t)\int_{\sigma-\etau}^{\sigma+\etau}\frac{P(\sigma',t)-P(\sigma,t)}{|\sigma-\sigma'|^{\mu+1}}\mathrm{d}\sigma'
    \nonumber\\ &&{}-\frac{\theta(|\sigma|-\sigma_c)}{\pltime}P(\sigma,t)+\Gamma(t)\delta(\sigma)
 \end{eqnarray}
where we have defined the yield rate
\begin{equation}
    \Gamma(t)=\frac{1}{\pltime}\int_{-\infty}^{\infty}\theta(|\sigma|-\sigma_c)P(\sigma,t)\mathrm{d}\sigma
\end{equation}
In the following we will generally consider the dimensionless form of (\ref{pde}), setting the threshold stress $\sigma_c=1$ and the plastic timescale $\pltime=1$.

Equation (\ref{pde}) is our desired time-dependent mean-field model for the elastoplastic dynamics of amorphous solids. It generalizes the model originally proposed by Lin and Wyart \cite{lin_mean-field_2016}, which was restricted to steady state scenarios. This model was described in terms of an accumulated plastic strain $\gamma^{\rm pl}$. It can be recovered from the general  formulation (\ref{pde})
by considering constant global stress $\langle\sigma\rangle$, i.e.\ a stress-controlled protocol. From (\ref{pde}) we have then $0=\partial_t \langle \sigma \rangle=G_0\dot{\gamma}-v \Gamma 
$ where the ``velocity'' $v$ relating shear rate and yield rate is the average stress of yielding sites, i.e.\ the average  over the distribution $\theta(|\sigma|-1)P(\sigma,t)/\Gamma(t)$. Using then that the plastic strain increments with each yield event so that $\dot\gamma^{\rm pl}=\Gamma$ 
one obtains equation (8) in Ref.~\onlinecite{lin_mean-field_2016} for $P(x,\gamma^{\rm pl})$, where $x=1-\sigma$.

The well-studied HL model \cite{hebraud_mode-coupling_1998} may also be derived as a limiting case of (\ref{pde}): it corresponds to the limit $\mu \rightarrow 2$, taken from below. To see this one can use a Kramers-Moyal expansion to express the convolution with a power law kernel in the second line of (\ref{pde}) as an infinite series of even-order derivatives $(\partial/\partial\sigma)^{2n}P$. The prefactor of the diffusive term ($n=1$) then works out as $\alpha_{\rm eff}\Gamma$ with
\begin{equation}\label{alpha_eff}
	\alpha_{\rm eff}=\frac{A}{2-\mu}\etau^{2-\mu}=\frac{A}{2-\mu}\left(\frac{2A}{\mu}\right)^{2/\mu-1}
\end{equation}
Keeping now $\alpha_{\rm eff}$ fixed while taking
$\mu \rightarrow 2^{-}$, one finds that $A\sim 2-\mu$ to leading order, with the consequence that all terms involving higher-order derivatives of $P$ ($(\partial/\partial\sigma)^4P$ etc) become negligible (see 
Appendix~\ref{app:model_construction} for details). The master equation then becomes that of the 
HL model~\cite{hebraud_mode-coupling_1998} 
\begin{eqnarray}\label{hl_equation}
\frac{\partial P(\sigma,t) }{\partial t} &=&-G_0 \dot{\gamma}\frac{\partial P}{\partial\sigma}
\label{HL}\\
&&{}+\alpha \Gamma(t)\frac{\partial^2 P}{\partial \sigma^2}-\theta(|\sigma|-1)P+\Gamma(t)\delta(\sigma)
\nonumber
\end{eqnarray}
where $\Gamma(t)$ is the yield rate as defined above while the stress propagation now takes the form of  Brownian motion with diffusion constant $D(t)=\alpha \Gamma(t)$. Note that for $\mu<2$ the relation (\ref{alpha_eff}) can be used to approximate our model (\ref{pde}) by an effective HL model, which as we will see below gives a reasonable qualitative account of the phase diagram.

\section{\label{sec:phase_diagram}Phase diagram}

The HL model \cite{hebraud_mode-coupling_1998} was shown, despite its simplicity, to display a \textit{dynamical arrest transition} as the coupling  $\alpha$ is varied. In particular, for $\alpha>\alpha_c=1/2$ (in dimensionless form) there exists a steady state distribution $\ps(\sigma)$ with nonzero yield rate $\Gamma$ even in the absence of external shear, and the low shear rate rheology is that of a Newtonian liquid. In contrast, below the critical value $\alpha_c$ only frozen steady states with $\Gamma=0$ exist and the system exhibits a finite yield stress, hence the description of it as being \textit{arrested}.

\begin{figure}
\includegraphics[scale=0.5]{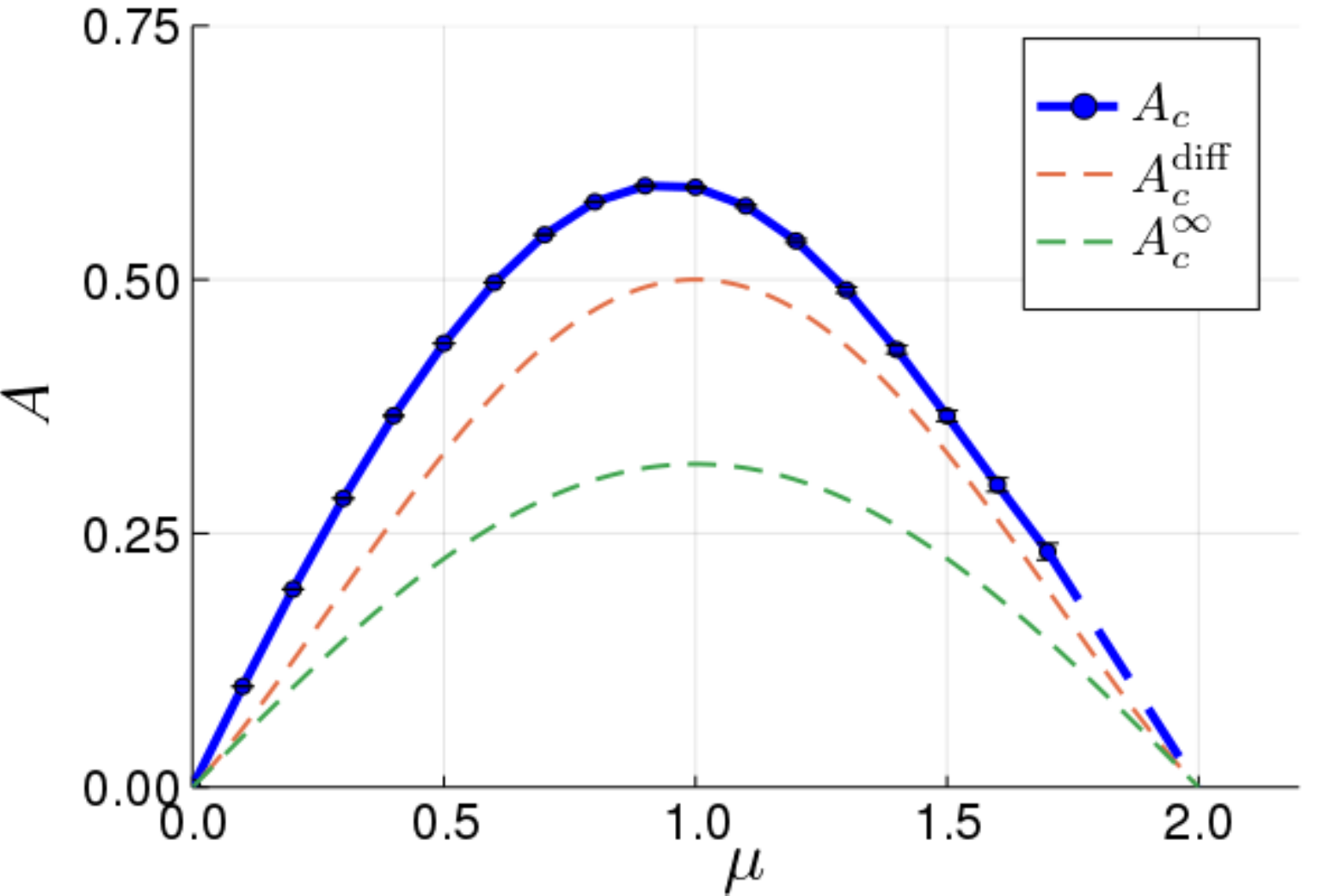}
\caption{\label{fig:phase_diagram}Phase diagram in the $A-\mu$ plane. The numerically exact $A_c(\mu)$ in blue separates the liquid regime (above) from the amorphous solid (below). Due to the increase in numerical error as $\mu \rightarrow 2$ (see Appendix~\ref{app:numerical_Stanley}), we show the last segment as an interpolation (blue dashed line). In addition, we show for comparison the diffusive (orange) and infinite cut-off (green) approximations to $A_c(\mu)$. }
\end{figure}

In the present model we expect a similar transition at a critical coupling $A_c$, and working out the corresponding phase diagram will be necessary to identify the correct parameter regime for studying the aging behaviour. In order to identify the critical value $A_c(\mu)$, which will also depend on the exponent $\mu$, it will be useful to consider the steady state version of (\ref{pde}) divided by $\Gamma$:
\begin{equation}\label{rescaled_pde}
A \int_{\sigma-\etau}^{\sigma+\etau}\frac{\ps(\sigma')-\ps(\sigma)}{|\sigma-\sigma'|^{\mu+1}}\mathrm{d}\sigma' +\delta(\sigma)-\frac{\theta(|\sigma|-1)}{\Gamma}\ps=0
\end{equation}
In the limit where $A \rightarrow A_c$ and correspondingly $\Gamma \rightarrow 0$, the final yielding term may be replaced by an absorbing boundary condition enforcing $\ps(\sigma)=0$ for $|\sigma|>1$. The critical boundary may then be computed numerically (for details see Appendix~\ref{app:numerical_Stanley}) giving the phase diagram shown in Figure~\ref{fig:phase_diagram}. The resulting curve $A_c(\mu)$ is bell-shaped, with a peak at $\mu \simeq 1$.

Also shown is a ``diffusive approximation'' $A_c^{\rm{diff}}$, which approximates the power-law noise for $\mu<2$ with Gaussian noise of the same variance. This is obtained by equating the $\alpha_{\rm eff}$ defined earlier in (\ref{alpha_eff}) to $\alpha_c=1/2$. This approach reproduces the general features of the curve $A_c(\mu)$. It becomes exact as expected in the limit $\mu \rightarrow 2$, where our model approaches the HL model. Conversely the approximation becomes worse (as can be seen by plotting e.g.\ the ratio $A_c/A_c^{\rm diff}$) as $\mu$ is decreased, especially in the region $\mu<1$ where the dynamics is increasingly dominated by large stress kicks. Interestingly, the approximation $A_c^{\rm diff}$ lies consistently below the true $A_c (\mu)$, indicating that the power-law noise is actually less efficient in ``liquifying'' the system than  Gaussian stress kicks with the same variance. Intuitively this can be rationalized from the large heterogeneity of power-law mechanical noise, where the variance is dominated by the large stress kicks while most kicks are in fact negligibly small. In studying the aging dynamics below we will find a similar effect, with the system aging towards an arrested state faster as $\mu$ is decreased.

We can develop a second approximation for $A_c(\mu)$ by exploiting a reinterpretation of (\ref{rescaled_pde}) as the steady state condition for what is known as a L\'evy flight~\cite{dubkov_levy_2008} with absorbing boundaries. 
Indeed, if we think of the local stress $\sigma$ as the position coordinate of an effective particle then this particle ``diffuses'' in power-law distributed steps, i.e.\ subject to L\'evy noise. This is the defining property of a L\'evy flight.
The reinjection term $\delta (\sigma)$ effectively sets the initial condition $\sigma=0$ of the particle, from which it executes its L\'evy flight until it hits the absorbing region $|\sigma|>1$.
Calling the resulting time-decaying distribution $g(\sigma,t)$, the problem may be solved by separation of variables, writing
\begin{equation}
    g(\sigma,t)=\sum \limits_{k}\psi_k (\sigma)e^{-t/\tau_k}\int_{-1}^{1}\psi_k(y)g(y,0)\mathrm{d}y
\end{equation}
in terms of the eigenmodes $\psi_k(\sigma)$ and their decay times $\tau_k$. These are related by the eigenvalue equation $A\mathcal{L}\psi_k (\sigma)=-\frac{1}{\tau_k}\psi_k(\sigma)$, where $\mathcal{L}$ denotes the propagator
\begin{equation}
\mathcal{L}\psi (\sigma)=\int_{\sigma-\etau}^{\sigma+\etau}\frac{\psi(\sigma')-\psi(\sigma)}{|\sigma-\sigma'|^{\mu+1}}\mathrm{d}\sigma'
\end{equation}
The eigenfunctions $\psi_k (\sigma)$ scale near the boundaries as~\cite{zoia_fractional_2007} $\psi_k (\sigma) \sim (1-|\sigma|)^{\mu/2}$. This singular behaviour will be picked up by the \textit{critical distribution}, defined as the limit of the steady state $P_c(\sigma)=\lim \limits_{A \rightarrow A_c}\ps(\sigma)$ as the critical coupling is approached (see Section~\ref{sec:scaling}).\footnote{ The exponent $\mu/2$ corresponds to the \textit{pseudogap} exponent in \cite{lin_mean-field_2016}}

Continuing with the L\'evy flight argument above, the survival probability of the effective particle at time $t$ may be written as $S(t)=\int_{-1}^{1}g(\sigma,t) \mathrm{d}\sigma$, and from this in turn we may derive the mean first passage time $\mfpt=\int_{0}^{\infty} S(t')\mathrm{d}t'$ until the particle is absorbed. To obtain a normalized steady state distribution $P_c(\sigma)$ at $A=A_c$, this mean lifetime needs to balance the reinjection which occurs at rate $1$, so that
\begin{equation}\label{MFPT_one}
    \mfpt=1
\end{equation}
This condition then implicitly determines the value of $A_c$.
In the HL model, one can follow the same argument and write the equivalent of equation (\ref{rescaled_pde}). Applying the well-known result for a particle diffusing with unit variance Brownian noise in a box, one finds $\mfpt (\alpha)=1/(2\alpha)$. With the condition (\ref{MFPT_one}) this then directly gives the critical coupling $\alpha_c=1/2$.

In our model with cut-off L\'{e}vy noise the mean lifetime cannot be obtained analytically, and $A_c(\mu)$ must be determined by finding $\tau_{\rm FP}$ numerically for a range of $A$ and solving for the value of $A$ -- the ``L\'{e}vy flight intensity'' -- where (\ref{MFPT_one}) is satisfied. In the absence of a cutoff, however, there is an analytical expression for the mean first passage time~\cite{buldyrev_average_2001}, which applied to our model gives
\begin{equation}
    A^\infty_c(\mu)
    =\frac{1}{\pi}\sin\left(\frac{\mu \pi}{2}\right)
\end{equation}
This is also included in Figure~\ref{fig:phase_diagram} for comparison. It shows again the same bell-shaped form; however, as one would expect it lies significantly below the true $A_c(\mu)$: without a cutoff the overall noise intensity is higher so the system stays liquid down to lower $A$.

\section{\label{sec:boundary_layer}Boundary layer equation}

To understand the scaling of the activity in steady state above the dynamical arrest transition, as well as for our later analysis of the aging behaviour, we introduce here a boundary layer framework. This is inspired by Ref.~\onlinecite{sollich_aging_2017}, where the HL model is studied in the same spirit.

It will be useful to write the local stress  distribution $P(\sigma,t)$ in terms of the yield rate as $P(\sigma,\Gamma)$. In the steady state the yield rate $\Gamma$ will be constant, whereas during the aging we expect it to decay in time. In either case we will consider an expansion of the master equation (\ref{pde}) for $\Gamma \ll 1$.

To motivate the approach below, consider the aging behaviour of $P(\sigma,t)$. Without shear to load local sites elastically, one expects that at long times the stress dynamics (\ref{pde}) will be dominated by local stress with values around the yield threshold $\sigma=\pm 1$, which we refer to as the boundary layer. To estimate the thickness of this layer we note that once the local stress at a site crosses the threshold it takes a typical time $\pltime$ to be reset to zero by a yield event. In this time it receives a number of kicks of order $\Gamma \pltime$, which in dimensionless units is just $\Gamma$. The typical stress changes occurring during this time are given by the Hurst exponent~\cite{dubkov_levy_2008} $H=1/\mu$, so that we expect the width of the boundary layer to scale as $\sigma\mp 1\sim\Gamma^{H}=\Gamma^{1/\mu}$. (The two signs relate to yielding at $+\sigma_c=+1$ and $-\sigma_c=-1$, respectively.)

\begin{figure}[th]
\includegraphics[scale=0.5]{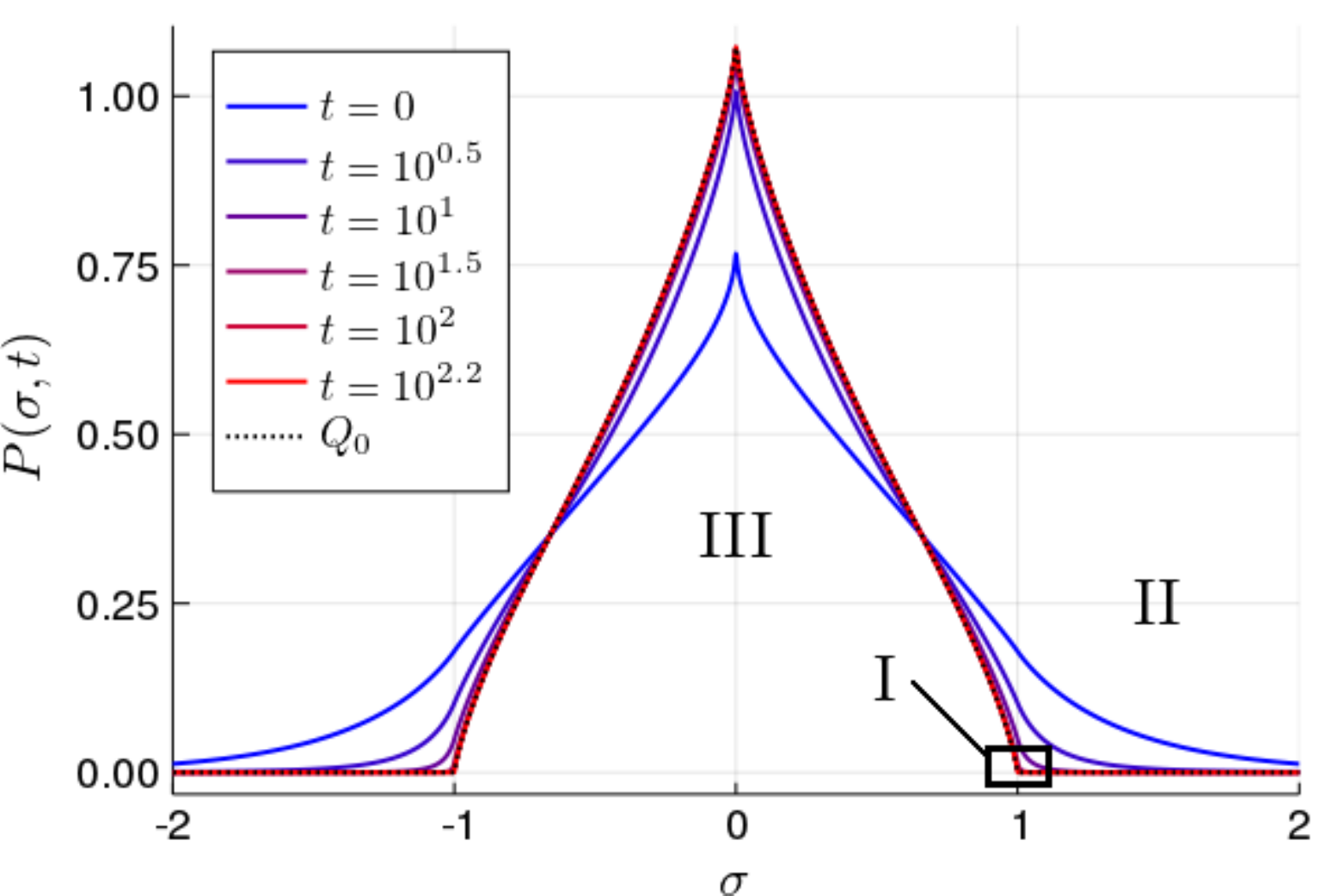}
\caption{\label{fig:three_regions_marked}Distribution $P(\sigma,t)$ at increasing (from blue to red) times $t$  during aging, obtained by numerical solution of (\ref{pde}), using as parameters $\mu=1.7$, $A=0.15$ starting from the reference steady state (see Sec.~\ref{sec:aging}). Also shown is the frozen-in distribution $Q_0(\sigma)$ (dotted line lying almost on top of distribution for $t=10^{2.2}$).}
\end{figure}

\begin{figure}[h]
\includegraphics[scale=0.5]{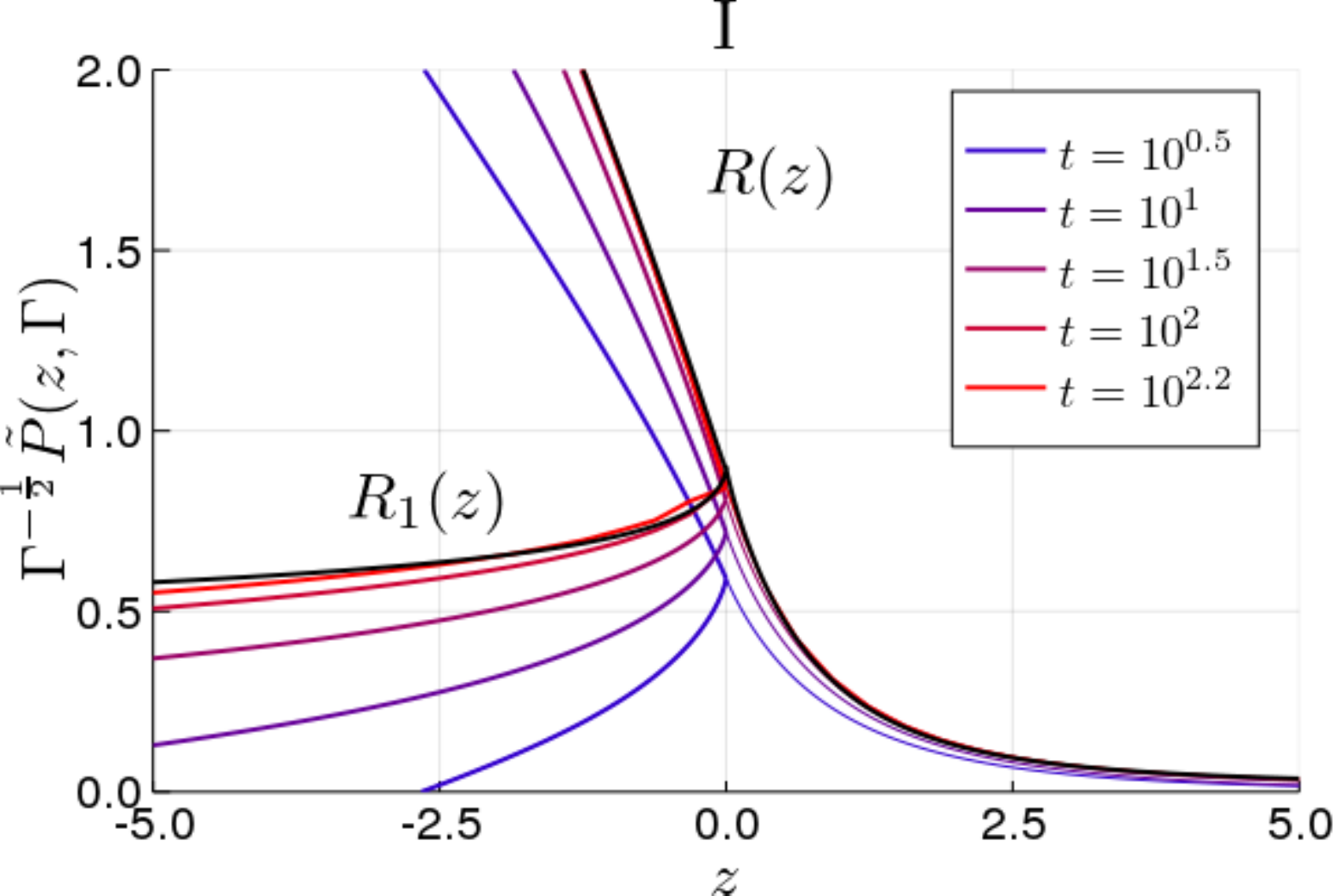}
\caption{\label{fig:first_region_BL}Boundary layer (zoom on region I of Fig.~\ref{fig:three_regions_marked}). In the upper curves we show $\Gamma^{-1/2}\tilde{P}(z,\Gamma)$, while in the lower curves we show $\Gamma^{-1/2}\tilde{P}(z,\Gamma)-\E (-z)^{\mu/2}\theta(-z)$, with $\E$ fitted from the frozen-in distribution $Q_0(\sigma)$; recall that $z=\Gamma^{-1/\mu}(\sigma-1)$. As $t$ increases and $\Gamma\to 0$, the lower curves approach $R_1(z)$, which is obtained by solving  (\ref{BL_eq}) numerically (see App.~\ref{app:numerical_BL}), while the upper curves collapse onto $R(z)$ from (\ref{Rz_split}).}
\end{figure}

\begin{figure}[h]
\includegraphics[scale=0.5]{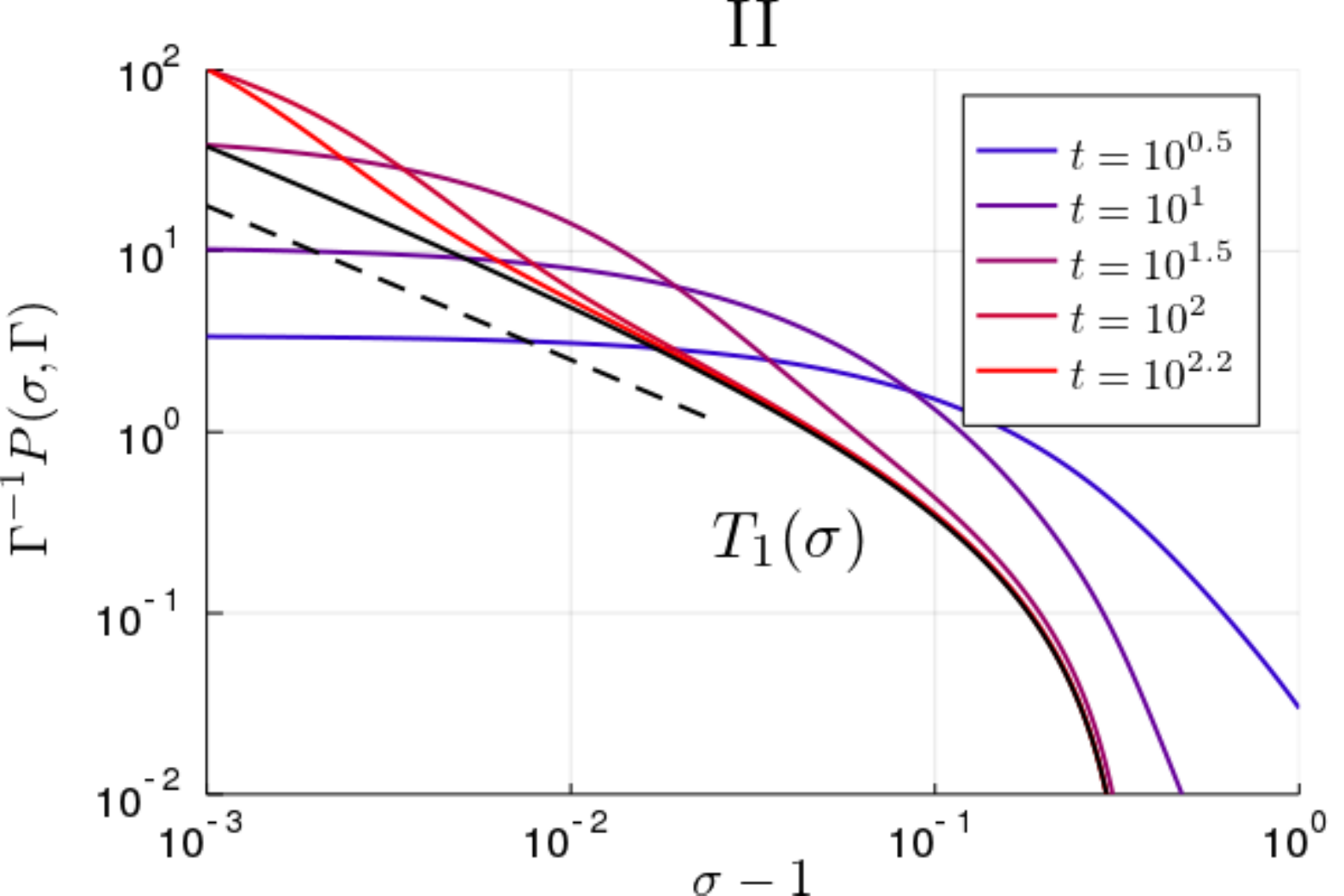}
\caption{\label{fig:second_region_exterior}External region (II) from Fig.~\ref{fig:three_regions_marked}. We show $\Gamma^{-1}P(\sigma,\Gamma)$, which for long times collapses onto $T_1(\sigma)$ (Eq.~\ref{T}). Also shown (dashed line) is the expected power-law behaviour $(\sigma-1)^{-\mu/2}$ for $\sigma-1\ll 1$ (\ref{T_approx}). }
\end{figure}

\begin{figure}[b]
\includegraphics[scale=0.5]{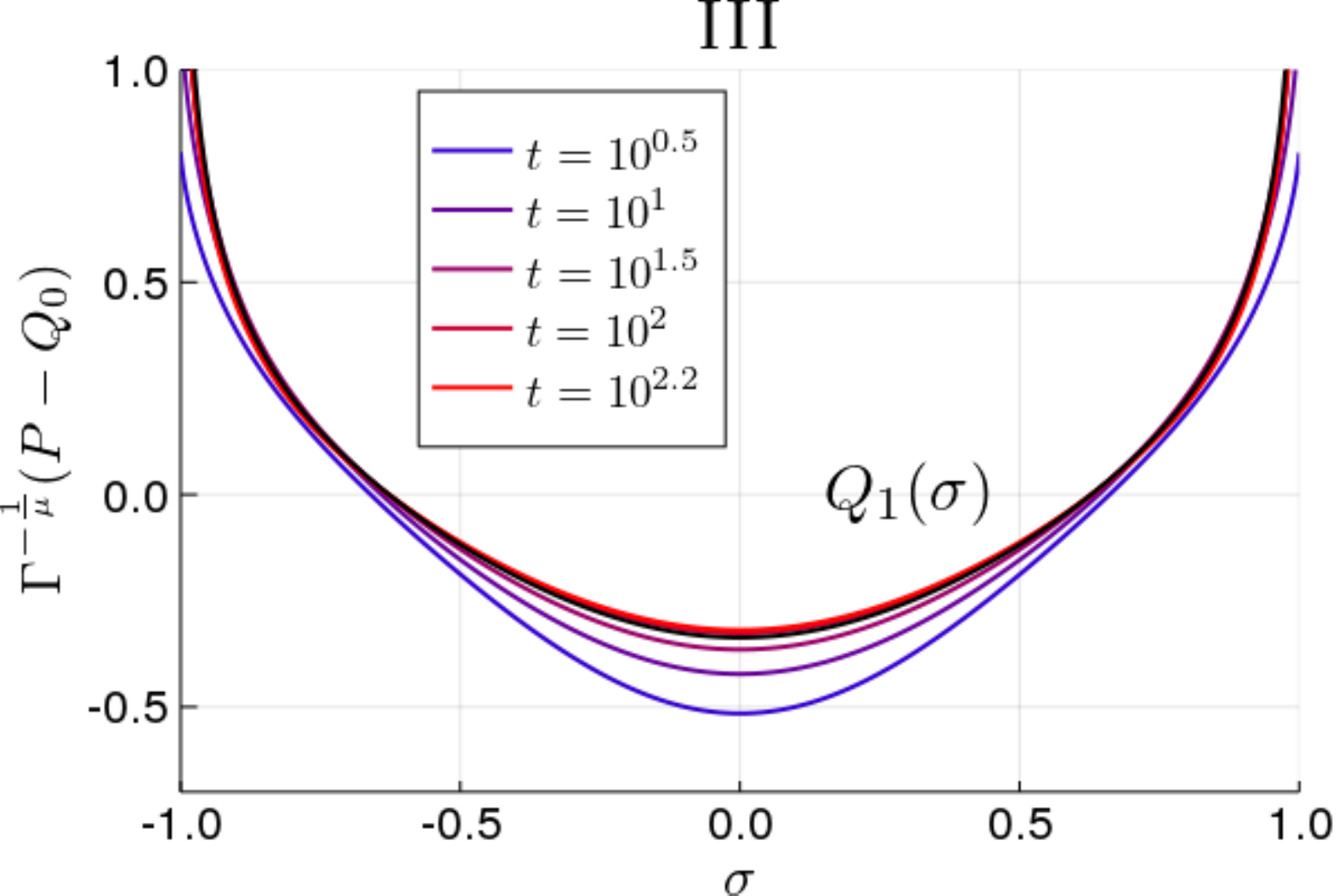}
\caption{\label{fig:third_region_interior}Interior region (III) from Fig.~\ref{fig:three_regions_marked}. We show $\Gamma^{-1/\mu}(P-Q_0)$, which at later times collapses onto the analytical prediction for the leading order interior function $Q_1(\sigma)$ from (\ref{Q1}).}
\end{figure}

To incorporate the presence of the boundary layer into our analysis we need to consider the behaviour of $P(\sigma,\Gamma)$ separately in the interior (sub-threshold) region $|\sigma|<1$, the exterior region $|\sigma|>1$ and the boundary layer. We consider symmetric distributions, $P(\sigma,\Gamma)=P(-\sigma,\Gamma)$, which in an unsheared steady state is automatic while for the aging dynamics it only requires a symmetric initial stress distribution.  
To connect the interior or exterior part of the distribution with the boundary layer we introduce a parameter $\epsilon$ such that $\Gamma^{1/\mu}\ll\epsilon \ll 1$. This then allows us to split the ansatz for $P(\sigma,\Gamma)$ into three different regions 
as sketched in Figure~\ref{fig:three_regions_marked} and verified from numerical simulation data in Figures \ref{fig:first_region_BL}, \ref{fig:second_region_exterior} and \ref{fig:third_region_interior}:
\begin{itemize}
\item In the interior region $|\sigma|<1-\epsilon$ (region III in Fig.~\ref{fig:three_regions_marked}) we  write $P(\sigma,\Gamma)=Q_0(\sigma)+\Gamma^{a}Q_1(\sigma)$, where $Q_0(\sigma)=\lim \limits_{\Gamma\rightarrow 0}Q(\sigma,\Gamma)$  and $Q_1$ is the leading order correction for small $\Gamma$. We will refer to $Q_0(\sigma)$ as the \textit{frozen-in} distribution, and it will present the $(1-|\sigma|)^{\mu/2}$ singularity discussed in Section~\ref{sec:phase_diagram}. In the scaling analysis for the steady state it will correspond to the critical distribution $P_c(\sigma)$. 
\item The exterior tail for $\sigma>1+\epsilon$ (region II in Fig.~\ref{fig:three_regions_marked}), which is symmetrically related to the left tail at $\sigma<-1-\epsilon$. In the exterior region we write the distribution as $P(\sigma,\Gamma)=\Gamma^{b}T_1(\sigma)$, where $b$ is the exponent of the leading order term.
\item In the boundary layer region (region I in Fig.~\ref{fig:three_regions_marked}), for  $1-\epsilon<\sigma<1+\epsilon$ (and similarly $-1-\epsilon<\sigma<-1+\epsilon$), we write the scaling function in terms of a rescaled stress variable $z=\Gamma^{-1/\mu}(\sigma-1)$ that from our argument above should be of order unity. In the boundary layer we therefore write  $P(\sigma,\Gamma)\equiv 	\tilde{P}(z,\Gamma)$ with the form of $	\tilde{P}(z,\Gamma)$ to be determined below.
\end{itemize}

The task is now to substitute the above ansatz into the master equation (\ref{pde}), separately for $\sigma$ in the three different regions: interior, boundary layer and exterior. We limit the discussion here to the boundary layer, which will be the most important for the physics, and leave further details for Appendix~\ref{app:scaling}.

In the boundary layer there will be a contribution from the frozen-in distribution. On the scale of the boundary layer this will be given by the singular behaviour $Q_0(\sigma)\simeq \E (1-\sigma)^{\mu/2}$, which written in terms of $z$ is $\Gamma^{1/2}\E (-z)^{\mu/2}$ with $z<0$. We deduce that $\tilde{P}(z,\Gamma)$ may be split into a frozen part and a correction, which again we write at leading order:
\begin{equation}\label{Split_frozen}
	\tilde{P}(z,\Gamma)=\Gamma^{1/2}\E(-z)^{\mu/2}\theta(-z)+\Gamma^c R_1(z)
\end{equation}
The exponent $c$ can be determined by matching the boundary layer function to $P(\sigma,\Gamma)$ in the exterior region, which decays as $(\sigma-1)^{-\mu/2}$ (see Appendix~\ref{app:scaling_external}), with the result $c=1/2$ so that the two exponents in (\ref{Split_frozen}) are equal and we can write (as before to leading order in $\Gamma$)
\begin{eqnarray}
    \tilde{P}(z,\Gamma)&=&\Gamma^{1/2}R(z)\\
    R(z) &=& \E(-z)^{\mu/2}\theta(-z)+R_1(z)\label{Rz_split}
\end{eqnarray}
From the master equation in the boundary layer region we can obtain an equation for the unknown scaling function $R_1(z)$, which we refer to as the \textit{boundary layer equation}. Without delving into details, three aspects are worth pointing out.

Firstly, in the boundary layer, cut-off effects disappear and we are left with pure power law behaviours. This is due to the rescaling introduced. Considering for example the L\'{e}vy propagation on $R_1(z)$, we have that 
    \begin{multline}\label{R_1(z)}
   \Gamma^{1/2}\int_{1-\epsilon}^{1+\epsilon}\frac{R_1(\Gamma^{-\frac{1}{\mu}}(\sigma'-1))-R_1(\Gamma^{-\frac{1}{\mu}}(\sigma-1))}{|\sigma-\sigma'|^{\mu+1}}\mathrm{d}\sigma'=\\\Gamma^{-1/2}\int_{-\epsilon \Gamma^{-\frac{1}{\mu}}}^{\epsilon \Gamma^{-\frac{1}{\mu}}}\frac{R_1(z')-R_1(z)}{|z-z'|^{\mu+1}}\mathrm{d}z'
    \end{multline}
We can then take the limits $\pm \epsilon \Gamma^{-1/\mu}$ of the integral to infinity because of the way we have defined the boundary layer, with $\Gamma^{1/\mu}\ll\epsilon$ and therefore $\epsilon\Gamma^{-1/\mu}\gg 1$.
    
A second important aspect of the boundary layer equation is that the interior and exterior scaling functions do not make any leading order contributions within the boundary layer; they only appear indirectly through the condition that the tails of $R_1(z)$ must match the behaviour of $T_1(\sigma)$ and $Q_1(\sigma)$ respectively as $\sigma \rightarrow 1$, as required  for continuity of $P(\sigma,\Gamma)$.

Finally, the terms arising from the time derivative $\partial_t$, which are from the beginning absent in the steady state case, will also be irrelevant in the aging because $\Gamma(t)$ will vary sufficiently slowly (see Appendix~\ref{app:scaling_contributions}).

Overall, the \textit{boundary layer equation} takes the form
\begin{equation}\label{BL_eq}
    A\int_{-\infty}^{\infty}\frac{R_1(z')-R_1(z)}{|z-z'|^{\mu+1}}\mathrm{d}z'+AS(z)-\theta(z)R_1(z)=0
\end{equation}
The second term is a source $S(z)$, which arises from applying the L\'evy propagator to the frozen part of the stress distribution. Explicitly it reads
\begin{equation}\label{S(z)}
		S(z)= \E \,\betaf\!\left (\frac{\mu}{2},1+\frac{\mu}{2}\right)z^{-\mu/2}, \quad z>0
\end{equation}
where $\rm{B}$ denotes the Beta function. Physically, then, the boundary layer equation describes how sites can have their local stress increased into the unstable region ($z>0$) by yield events elsewhere; this is the source term $AS(z)$. These sites can then yield as indicated by the last term in (\ref{BL_eq}), or their stress may change due to further stress kicks (first term). The equation then simply states that these effects must balance in a stationary or slowly aging system.

The boundary layer equation (\ref{BL_eq}) admits different asymptotic solutions in the regimes $1<\mu<2$, $\mu=1$ and $\mu<1$. This will be the key to the distinct scaling and aging behaviours in the different regimes. We defer a detailed analysis (in particular for the case $\mu=1$) to Appendix~\ref{app:scaling_mu_1} and only outline the main properties here.

Splitting $R_1(z)$ into $\Rext(z)$ for $z>0$ and  $\Rint(z)$ for $z<0$, we focus on the asymptotic forms of these functions for $|z|\gg 1$. As indicated by the superscripts, $\Rext(z)$ will have to match up with $P(\sigma,\Gamma)$ in the exterior region, while $\Rint(z)$ will need to do so with the interior 
function.

It is straightforward to deduce from equation (\ref{BL_eq}) that the exterior asymptote must balance the form of the source, so that
\begin{equation}
    \Rext(z)=C_{\rm{ext}} z^{-\mu/2}\quad {\rm for} \quad z \gg 1
\end{equation}
For the interior asymptotic behaviour we assume a power law form $\Rint(z)=C_{\rm{int}}|z|^{-\abso}$ and find by appropriate rescaling of the integration variable that the boundary layer equation (\ref{BL_eq}) in the regime $|z|\gg 1$, $z<0$ becomes 
\begin{multline}\label{terms}
        \left(\int_{0}^{\infty}\frac{x^{-\abso}-1}{|x-1|^{\mu+1}}\mathrm{d}x-\frac{1}{\mu}\right)C_{\rm{int}}|z|^{-\abso-\mu}\\
        +\int_{0}^{\infty}\frac{\Rext(z')}{|z-z'|^{\mu+1}}\,\mathrm{d}z'=0
\end{multline}
There are then two possible solutions for the exponent $\abso$. In the regime $1<\mu<2$, one can show that the final integral term  may be neglected, and equation (\ref{terms}) can then be solved by using the property
\begin{equation}\label{two_behaviours}
    \int_{0}^{\infty}\frac{x^{-\abso}-1}{|x-1|^{\mu+1}}\mathrm{d}x=\frac{1}{\mu}\quad {\rm for}\quad \abso=-\frac{\mu}{2} \mbox{\ or\ } 1-\frac{\mu}{2}
\end{equation}
We choose the second value $\abso=1-\mu/2$ as we expect a decaying power law for $\Rint(z)$. This solution we refer to as the \textit{homogeneous} solution, given that it is obtained by neglecting the integral term, which represents a source arising from jumps from the unstable region $z'>0$. If this integral term were absent, the power law solution for $\Rint(z)$ would hold for $\forall z<0$, not just asymptotically; it would then correspond to one of the two alternative power law behaviours (whose exponents are the two values of $\abso$ in (\ref{two_behaviours})) of a L\'{e}vy flight in front of an absorbing boundary.

In the regime $\mu<1$ we find that we require an \textit{inhomogeneous} solution of (\ref{terms}), where all terms are kept. Here we take instead $\abso=\mu/2$, which with an appropriate choice of $C_{\rm{int}}$ allows us to cancel the source term from $z'>0$. Finally, in the marginal case $\mu=1$, where the exponent crosses over from $\abso=1-\mu/2$ to $\abso=\mu/2$, one can show (Appendix~\ref{app:scaling_mu_1}) that a superposition of both solutions gives the form $\Rint(z)=C_{\rm{int}}z^{-1/2}\ln{(|z|)}$, involving a log-correction to the power law. Summarizing, the  boundary layer function $R_1$ on the interior side has the asymptotic behaviour
\begin{equation}
    \Rint(z)\overset{|z|\gg 1}{\sim}
    \left\{
    \begin{array}{lr}
      |z|^{-(1-\mu/2)},& {\rm{for}} \quad 1<\mu<2 \\
      |z|^{-1/2}\ln(|z|),& {\rm{for}} \quad \mu=1 \\
      |z|^{-\mu/2},& {\rm{for}} \quad \mu<1
    \end{array}
    \right.
\end{equation}

The physical implications of these solutions will be discussed further below. In brief, these will arise because as explained above, the interior tail of the boundary layer function $\Rint(z)$ has to match the $\sigma \rightarrow 1^{-}$ behaviour of $Q_1(\sigma)$, and will therefore determine the scaling of the interior correction with $\Gamma$, given by the exponent $a$ defined above. Explicitly, for the case $1<\mu<2$ in the asymptotic $|z|\gg 1$ tail one has $\Gamma^{1/2}\Rint(z)\sim \Gamma^{1/2}z^{-(1-\mu/2)}$, which has to match with $Q_1(\sigma)\sim\Gamma^{a}(1-\sigma)^{-(1-\mu/2)}$, so that $a=1/\mu$. Arguing similarly for $\mu <1$, one finds $a=1$. For the case $\mu=1$, if we expand the asymptotic behaviour of $\Rint (z)$ in terms of $\sigma$ we find
\begin{equation}
    \Gamma^{1/2}\Rint(z)\sim\Gamma (1-\sigma)^{-1/2}\left[|\ln{(\Gamma)}|+\ln{(1-\sigma)}\right]
\end{equation}
The matching must take place for $1-\sigma \gtrsim \epsilon$, where $\epsilon$ was defined as $1\gg\epsilon\gg\Gamma^{1/\mu}$, so that the second term in the brackets may be neglected. Overall we then have that for $\sigma \rightarrow 1^{-}$ the interior function has to behave as $Q_1(\sigma)\sim \Gamma |\ln{(\Gamma)}|(1-\sigma)^{-1/2} $, so that $a=1$ with a logarithmic correction.

\section{\label{sec:scaling}Steady state scaling}

\begin{figure}
 \hspace{0.5cm}
\includegraphics[scale=0.45]{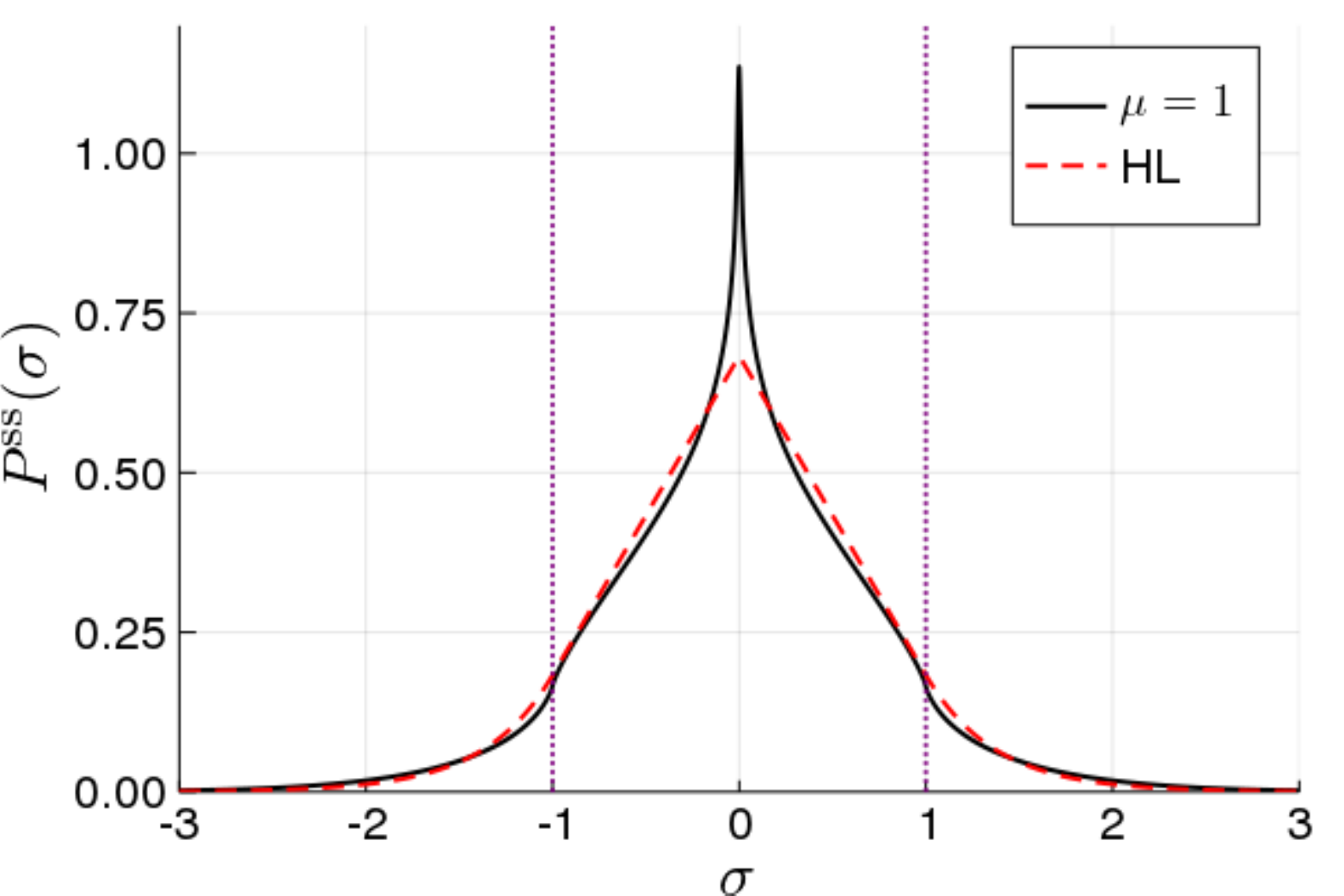}
\caption{\label{fig:steady_states} Steady state stress distribution in the liquid regime for $\mu=1$ and the HL model ($\mu\to 2$), for fixed yield rate $\Gamma=0.134$. Dotted lines mark the local yield thresholds $\sigma_c=\pm 1$. Whereas in the HL limit the form is a combination of exponentials and linear segments, for $\mu<2$ there is no simple analytical expression.}
\end{figure}

We now apply the approach above to deduce the critical scaling of the plastic rate above the transition. In the HL model (Eq.~\ref{HL}), the steady state distribution $\ps (\sigma)$ can be found analytically for any~\cite{agoritsas_relevance_2015,pituk_stability_2011} $\alpha >\alpha_c$. In Figure~\ref{fig:steady_states} we show the result 
for $\alpha=1$, together with the steady state in the $\mu=1$ model, with the same yield rate $\Gamma$. Although this value is well within the liquid regime, one may already note the differences between the two models: namely the straight line segments and exponential tails of the HL model become forms with no simple analytical expression.

For the HL model one can find an exact relation between $\alpha$ and $\Gamma$, which comes out of the normalisation condition on $\ps(\sigma)$. Writing $\alpha$ 
in terms of a rescaled distance to the transition $\tilde{\alpha}$, this relation reads:
\begin{equation}
 	\tilde{\alpha}\equiv\frac{ \alpha-\alpha_c}{\alpha_c}=2\left(\sqrt{\alpha \Gamma}+\alpha \Gamma \right)
\end{equation}
so that
\begin{equation}
\tilde\alpha = \mathcal{O}(\Gamma^{1/2})
\label{alpha_tilde_HL}
\end{equation}
for $\Gamma \ll 1$.

For the present model, on the other hand, we lack an analytical expression of the steady state stress distribution and its norm in terms of ($A$,$\Gamma$), and have to calculate the latter numerically in general. To find the critical scaling of $\Gamma$ above the transition analytically, we then proceed by considering a perturbation of the critical distribution $P_c(\sigma)$ defined in Section~\ref{sec:phase_diagram}. We defer the details to App.~\ref{app:scaling_ss}, where considering the steady state condition (\ref{rescaled_pde}), we show that one can again express the problem in terms of the boundary layer equation (\ref{BL_eq}). This leads to the following scalings
\begin{equation}\label{A_scaling}
    \tilde{A}\equiv\frac{A-A_c}{A_c}=
    \left\{
    \begin{array}{lr}
      \mathcal{O}(\Gamma^{1/\mu}),& {\rm{for}} \quad 1<\mu<2 \\
      \mathcal{O}(\Gamma \,|\!\ln({\Gamma})|),& {\rm{for}} \quad \mu=1 \\
      \mathcal{O}(\Gamma),& {\rm{for}} \quad \mu<1
    \end{array}
    \right.
\end{equation}
\begin{figure}
\includegraphics[scale=0.5]{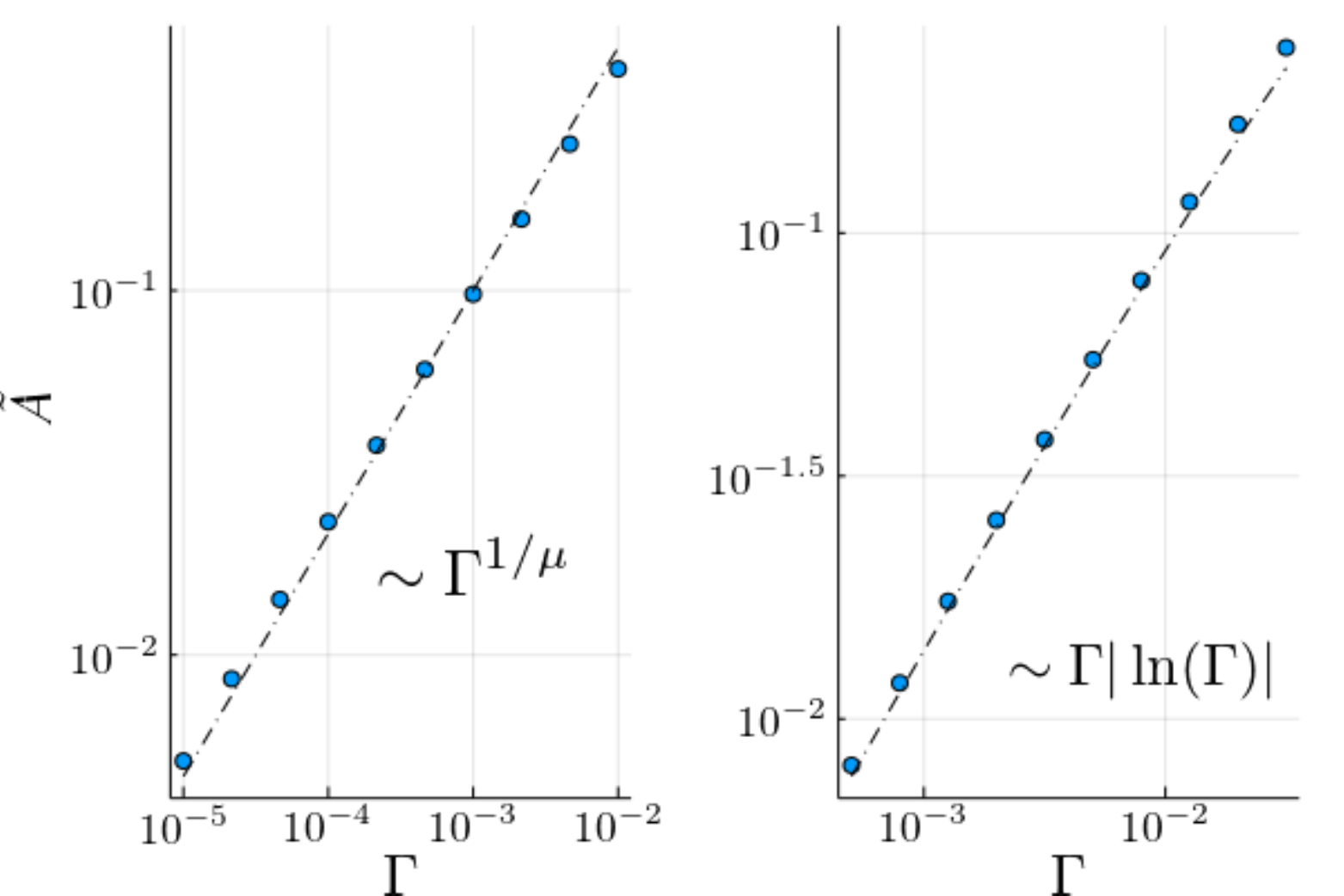}
\caption{\label{fig:scaling} Critical scaling of plastic activity, obtained numerically for $\mu=1.5$ (left) and $\mu=1$ (right). The dashed lines show the theoretical prediction according to (\ref{A_scaling}).}
\end{figure}

In Figure~\ref{fig:scaling} we show our numerical results (see Appendix~\ref{app:numerical_Stanley} for details) for the asymptotic scaling of $\tilde{A}$ with $\Gamma$. The results are in excellent agreement with the analytical prediction above (\ref{A_scaling}), both for $\mu=1.5$ and $\mu=1$. Note also that for $\mu\to 2$, our prediction recovers the scaling in the known HL case (\ref{alpha_tilde_HL}).

The analysis of the steady state regime also provides intuition for the more complicated aging behaviour that we study next. Following the interpretation of the dynamics as that of an effective particle diffusing from the origin, we show in App.~\ref{app:scaling_ss} that $\tilde{A}\sim \Delta \tau^{\rm{ext}}$, where $\Delta \tau^{\rm{ext}}$ is the extra time that a particle lives before yielding when $\Gamma>0$.
This extra time arises because yielding is no longer effectively instantaneous for nonzero $\Gamma$, and from (\ref{A_scaling}) it scales as $\Delta \tau^{\rm ext}\sim \tilde{A}\sim \Gamma^{1/\mu}$.
Now during the aging the effective diffusion process will take place at ever decreasing $\Gamma$. For $\Gamma \ll 1$ we then see that $\Delta \tau^{\rm ext}\sim \Gamma^{1/\mu}\ll \Gamma^{1/2}$ for $\mu<2$. The extra time the particle lives before yielding $\Delta \tau^{\rm ext}$ thus drops to zero faster for $\mu<2$, so that the system will not be able to sustain so many yield events and will age towards an arrested state faster than in the HL case. We will see this intuition confirmed in the analysis below.

\section{\label{sec:aging}Aging in the glassy regime}

 
In Section~\ref{sec:phase_diagram} the phase diagram of the model was presented, and we distinguished between a liquid and a glassy phase depending on the strength of the coupling $A$. In the liquid region $A>A_c$ there is a steady state distribution $\ps (\sigma)$ with a constant plastic activity $\Gamma >0$. As a representative for the liquid regime and a reference distribution for the following, we may take the $\ps(\sigma)$ with the yield rate corresponding to that of the HL model with $\alpha=1$ ($\Gamma=0.134$). This distribution (for $\mu=1$) is shown in Figure \ref{fig:steady_states}, along with its HL analogue. 
 
We are interested in the aging behaviour in the glassy regime $A<A_c$, where due to the absence of a steady state distribution with $\Gamma>0$ the plastic activity will decay as the system approaches some, potentially initial condition dependent, frozen-in stress distribution $Q_0 (\sigma)$. For the purposes of studying the slow decay of $\Gamma(t)$, we assume that the system starts in a configuration with unstable sites, which may be the result of an initial preparation such as stirring, and the system is then left to evolve at a value of the coupling $A<A_c$. To simplify the analytical study we focus here on symmetric initial distributions. We have considered as initial configuration with unstable sites mainly the above $\ps(\sigma)$, but also top hat and Gaussian distributions, with qualitatively identical results (data not shown). Asymmetric initial distributions could arise e.g.\ by pre-shear. Regardless of how the initial state is prepared, we study here the form of the asymptotic decay of $\Gamma(t)$, which we expect to be qualitatively unaffected by the details of the initial distribution~\footnote{Although we have not studied in detail the aging of asymmetric distributions in the current model, for the HL model~\cite{sollich_aging_2017} the same asymptotic decay and scalings are shown to hold also for the asymmetric case, under generic assumptions.}.

The aging behaviour considered here can be thought of as modelling the dissipative decay in activity in an athermal system, which is left to evolve in quiescent conditions after, for example, shear melting~\cite{agarwal_signatures_2019} or a sudden increase in density in thermosensitive core-shell microgel particles, whose size can be controlled by varying temperature~\cite{purnomo_glass_2008}. In the ensuing dynamics, given the athermal nature of the system rearrangements may only be triggered by events taking place elsewhere in the material. In a recent study of such relaxation in an athermal system \cite{chacko_slow_2019}, rearrangements are found to be reminiscent of the Eshelby events considered here.

 \begin{figure}[th]
\includegraphics[scale=0.6]{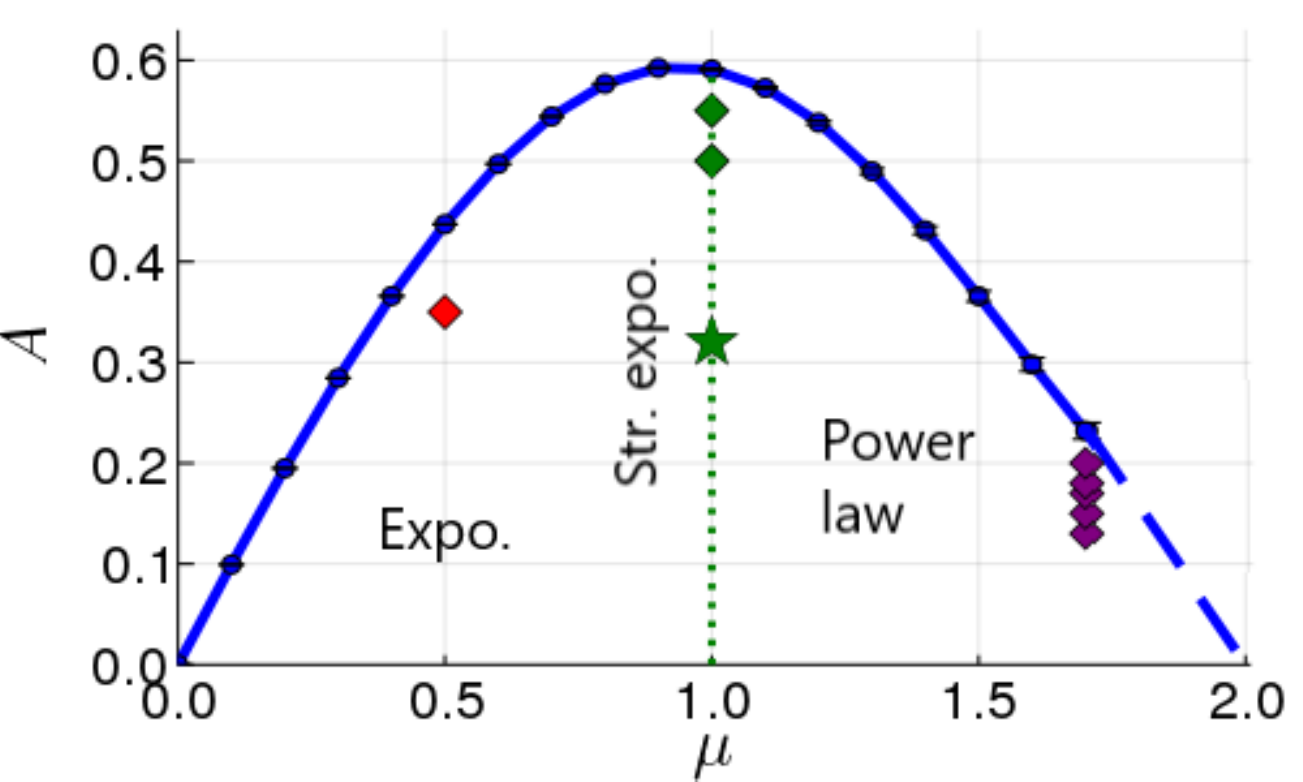}
\caption{\label{fig:aging_points}Location in the phase diagram of the points where numerical results on the aging behaviour are shown: power law region $(\mu=1.7,A=0.2$, 0.18, 0.17, 0.15 and 0.13, purple, see Figs.~\ref{fig:mu_17_gammas}, \ref{fig:logders}), stretched exponential line $(\mu=1,A=0.55$ and 0.5, green, see Figs.~\ref{fig:sqrt_expo}, \ref{fig:log_square}), exponential decay $(\mu=0.5, A=0.35,$ red, see Fig.~\ref{fig:gamma_mu_05}). The green star indicates the parameters of the lattice model $(\mu=1, A=0.32$, see Sec.~\ref{sec:lattice}). This last position should be taken as a qualitative indication only, given that the stress kick distribution in the lattice model is not exactly of the upper cutoff form (\ref{upper_cutoff}); this form affects the precise location of the phase boundary.}
\end{figure}

 \subsection{\label{subsec:level1}Aging for $1<\mu<2$}
 
 To study the aging behaviour, we turn again to the expansion in $\Gamma \ll 1$ presented in Section~\ref{sec:boundary_layer}. In contrast to the analysis of the steady state scaling, however, we will now have a rate of rearrangements that decreases in time. Nevertheless one may still consider the same boundary layer ansatz, now with $\Gamma(t)$.

We defer the detailed analysis to Appendix~\ref{app:scaling}. The most important step is establishing the equation of motion in the interior region. Following the general ansatz presented in Section~\ref{sec:boundary_layer}, the distribution is to leading order $P(\sigma,\Gamma)=Q_0(\sigma)+\Gamma^{1/\mu}Q_1(\sigma)$. The equation of motion in this region can then be written as
\begin{equation}
    \partial_t P=\frac{1}{\mu}\Gamma^{1/\mu-1}\dot{\Gamma}Q_1=A\Gamma \mathcal{L} \left( \Gamma^{1/\mu}Q_1 +Q_0 \right)+\Gamma\delta(\sigma)
    \label{aging_dP_dt}
\end{equation}
up to terms of higher order in $\Gamma$ that we omit. We may now distinguish between two cases. Firstly, in the critical case $A=A_c$ (critical aging) the combination $A\mathcal{L}Q_0+\delta(\sigma)$ vanishes because of the limiting (for $\Gamma\to 0$) steady state condition, so that the prefactors of the remaining terms in (\ref{aging_dP_dt}) involving $Q_1$ must be the same. This gives $\dot\Gamma\sim \Gamma^2$ and hence $\Gamma(t)\sim1/t$. This is independent of $\mu$ and so one would expect the same critical behaviour to hold also for $\mu\to 2$, i.e.\ for the HL model. This can indeed be shown, by extending the analysis in Ref.~\onlinecite{sollich_aging_2017}; we omit the details.

In the generic case $A<A_c$ the sum $A\mathcal{L} Q_0+\delta(\sigma)$ in (\ref{aging_dP_dt}) no longer vanishes. As this sum appears in (\ref{aging_dP_dt}) with prefactor $\Gamma$, in order to balance it and obtain a time-independent $Q_1$, we also need the left hand side to be of order $\Gamma$. This leads to the conclusion that $\dot{\Gamma}\sim \Gamma^{2-1/\mu}$, from which one deduces that $\Gamma(t)$ decays in time in a power-law fashion as
\begin{equation}\label{decay}
    \Gamma(t)=\prefaging t^{-\mu/(\mu-1)}
\end{equation}
Here the prefactor $\prefaging$ is related to the frozen-in distribution  $Q_0(\sigma)$, and like the latter is therefore  expected to be dependent on initial conditions. In the limit $\mu \rightarrow 2^{-}$, this prediction is consistent with the aging behaviour $\Gamma\sim t^{-2}$ that was found independently for the HL model \cite{sollich_aging_2017}.

To test the decay (\ref{decay}) numerically, we perform numerical tests by using a pseudospectral method on a discrete grid of $\sigma$-values (see Appendix~\ref{app:numerical_pseudospectral}) to solve the equation of motion (\ref{pde}) for $P(\sigma,t)$. We choose as initial condition the liquid steady state with $\Gamma=0.134$ (see above), and set $A$ to a value below the dynamical arrest transition.

\begin{figure}[th]
\includegraphics[scale=0.5]{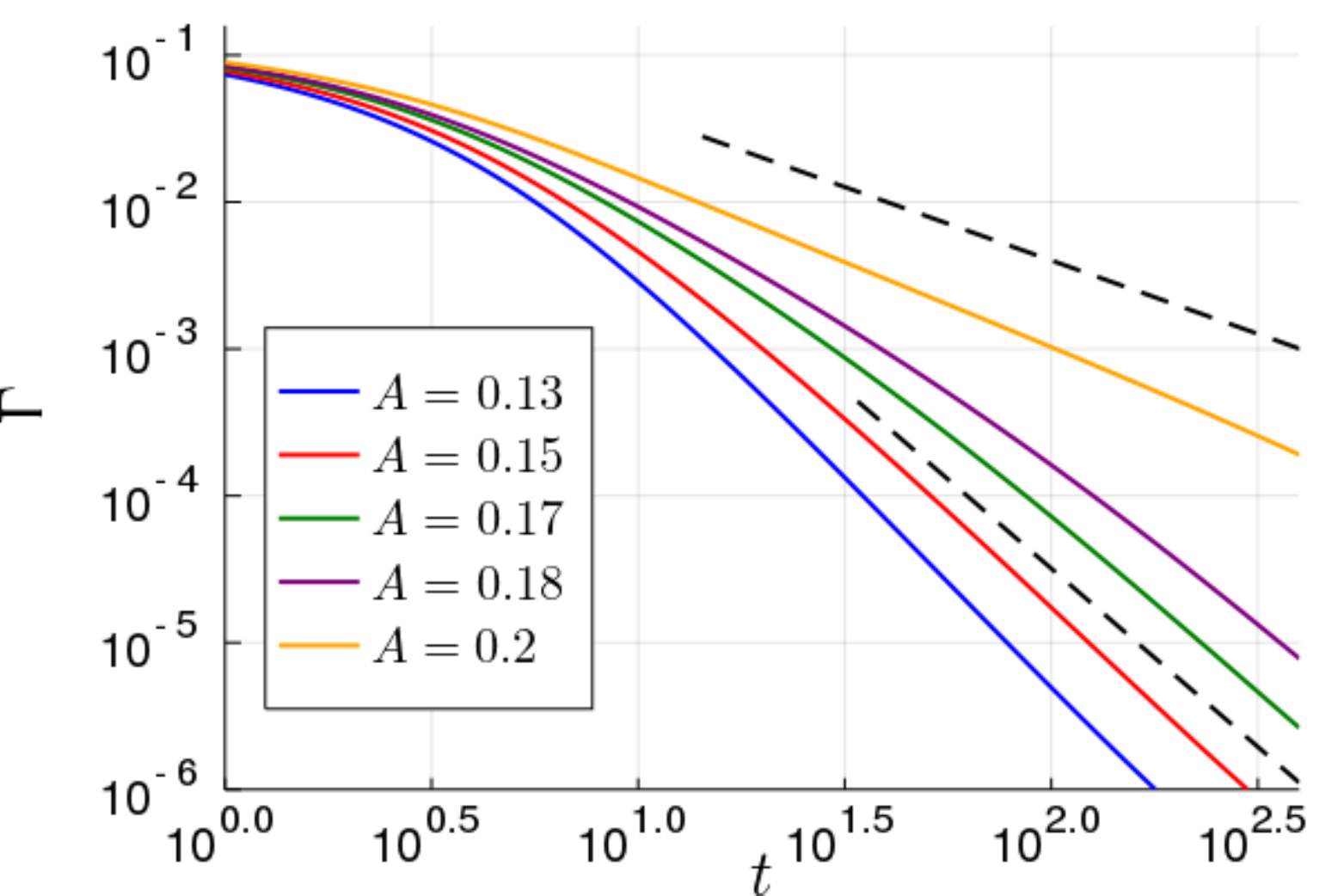}
\caption{\label{fig:mu_17_gammas}Decay of $\Gamma(t)$ for $\mu=1.7$, at five different values of $A$ below the dynamical arrest transition, starting from the liquid steady state with $\Gamma=0.134$. The lower dashed line shows the predicted power law asymptote $\Gamma\sim t^{-\mu/(\mu-1)}$, with $\mu/(\mu-1)\simeq2.428$. The upper dashed line shows the asymptotic power law at $A=A_c$, $\Gamma \sim t^{-1}$.}
\end{figure}

\begin{figure}[th]
\includegraphics[scale=0.48]{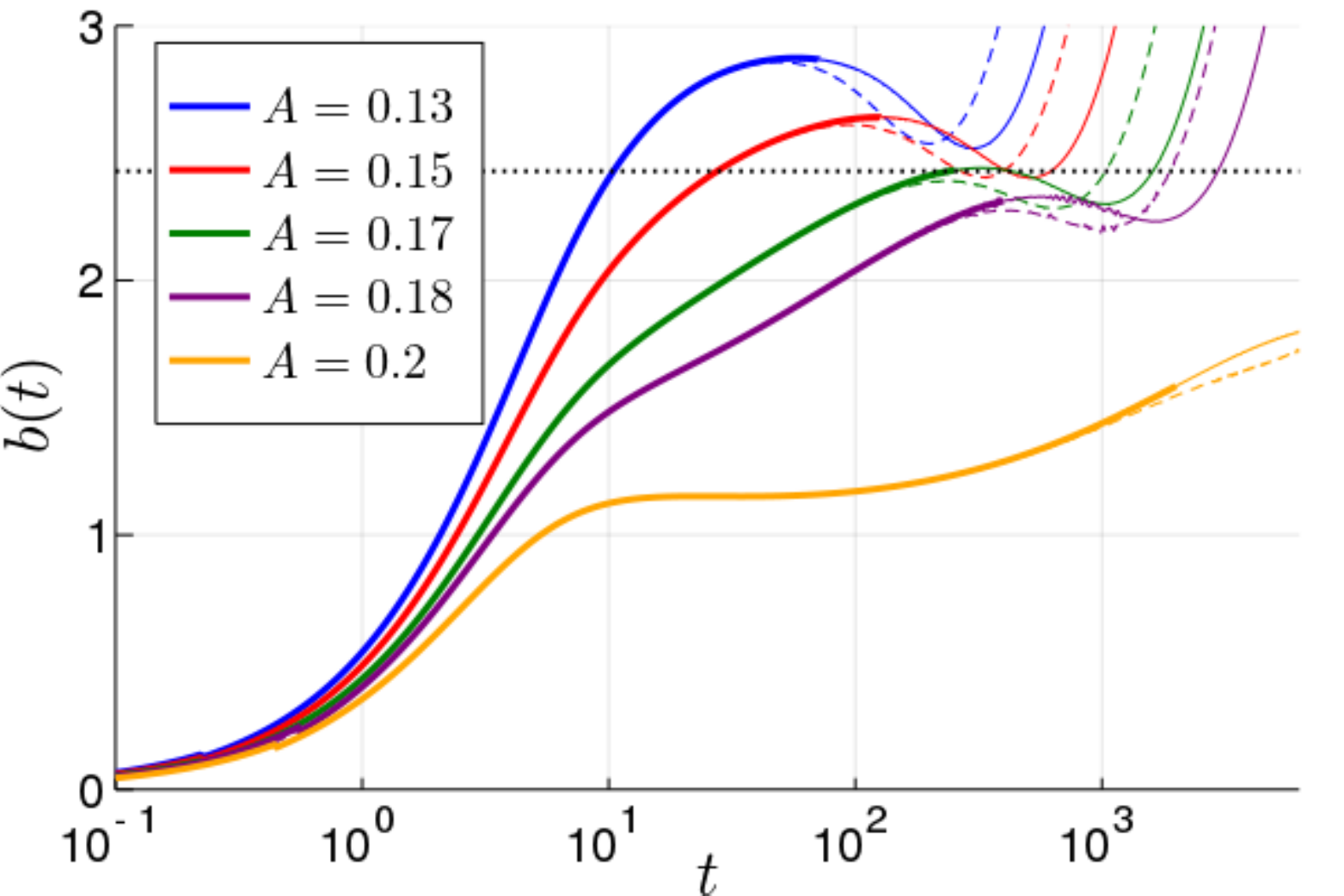}
\caption{\label{fig:logders}Evolution of the log derivative $b(t)$ of the yield rate, defined in the text, for the same runs as in Figure~\ref{fig:mu_17_gammas}. The four upper curves ($A=0.13,0.15,0.17,0.18$) converge towards the predicted exponent value $\mu/(\mu-1)\simeq2.428$ (dotted line). The highest value of $A=0.2$ is affected by the critical behaviour $\Gamma \sim t^{-1}$ at $A=A_c$. Dashed curves indicate runs on a coarser (by a factor of $2$) $\sigma$-grid; the limit of reliability of the numerical results is reached where these start to deviate (thinner lines).}
\end{figure}

The results are displayed in Figures \ref{fig:mu_17_gammas} and \ref{fig:logders}, where we choose $\mu=1.7$ throughout; Fig.~\ref{fig:aging_points} gives an overview of where the $(\mu,A)$-pairs are located in the phase diagram. In the first plot (Fig.~\ref{fig:mu_17_gammas}) we show the behaviour of $\Gamma(t)$ in log-log scale, for several values of $A$ below the transition, along with the corresponding predicted power law. We see a good agreement, although reaching the asymptotic behaviour is challenging. Numerically, this is due to the fact that one can no longer obtain reliable numerical data when the scale of the boundary layer $\Gamma^{1/\mu}$ becomes of the order of the discretization interval $d\sigma$. This problem is accentuated as $\mu$ is decreased, due to the scaling $\Gamma^{1/\mu}$ which implies a narrower boundary layer. Also, lower values of $A$ are more challenging, given that the yield rate decays more quickly and the above numerical limit is reached sooner. The limit can be seen in Fig.~\ref{fig:logders} as the point where results for different $\sigma$-discretizations start to differ (thin lines). 

Physically, the covergence to the power-law asymptote is affected by two crossover behaviours. To illustrate this, in the second plot (Fig.~\ref{fig:logders}) we show the log-derivative
\begin{equation}
    b(t)=-\frac{\rm{d}\ln (\Gamma)}{\rm{d}\ln (t)}
\end{equation}
This is an effective power law exponent that at long times should converge to the value $\mu/(\mu-1)$. On the one hand, for larger $A$ (see $A=0.2$ in Fig.~\ref{fig:logders}) closer to $A=A_c$ the convergence is affected by a transient where the system behaves as in the critical case, with the yield rate decaying as $\Gamma\sim t^{-1}$. For smaller $A$ (see $A=0.13$), on the other hand, $b(t)$ increases very rapidly in the transient. This is because here we approach the limiting behaviour for $A\rightarrow0$, where there is no stress redistribution and $\Gamma(t)$ decays as a pure exponential. This rapid increase in $b(t)$ at low $A$ is the reason why for $A=0.13$ and 0.15 in Fig.~\ref{fig:logders} the effective exponent converges to its limit from above, while for the higher values $A=0.17$ and 0.18 it does so from below.
 
 \subsection{\label{subsec:level2}Aging for $\mu=1$}
 
We next turn to the aging behaviour in the marginal case $\mu=1$. As outlined in Sec.~\ref{sec:boundary_layer}, in this case the \textit{homogeneous} and the \textit{inhomogeneous} solutions of the boundary layer equation merge, resulting in a log-correction to the asymptotic power law on the interior side of the boundary layer function $\Rint(z)\sim |z|^{-1/2}\ln(|z|)$. This now determines the form of the interior distribution, whose non-frozen part has to match $\Gamma^{1/2}R_1(z)$ for $\sigma\to 1$. As described in Section~\ref{sec:boundary_layer},
using that $z=(\sigma-1)/\Gamma$ for $\mu=1$ one has $\Gamma^{1/2}|z|^{-1/2}\ln(|z|) \simeq |\sigma-1|^{-1/2}\Gamma|\!\ln(\Gamma)|$ to leading order so that the distribution in the interior must take the form $P(\sigma,\Gamma)=Q_0(\sigma)+\Gamma |\!\ln (\Gamma)| Q_1(\sigma)$. Proceeding as in the case $\mu>1$, the time dependence of $\Gamma$ may now be derived from the equation of motion in the interior, where the time decay of the bulk of the probability distribution has to balance the jumps out to the unstable region, leading to
\begin{equation}
    \partial_t\left(-\Gamma \ln(\Gamma)\right)=-\prefexp \Gamma
\end{equation}
where $\prefexp>0$ is a constant. Solving this differential equation for $\Gamma(t)$ leads to
\begin{equation}\label{stretched}
    \Gamma(t){\sim}e^{-\sqrt{2\prefexp t}}
\end{equation}
In the long time regime, therefore, we obtain a {\em stretched exponential} behaviour. The prefactor of the power law ($\prefexp$) is initial condition dependent as it is again related to the frozen-in distribution $Q_0(\sigma)$.

\begin{figure}
\includegraphics[scale=0.48]{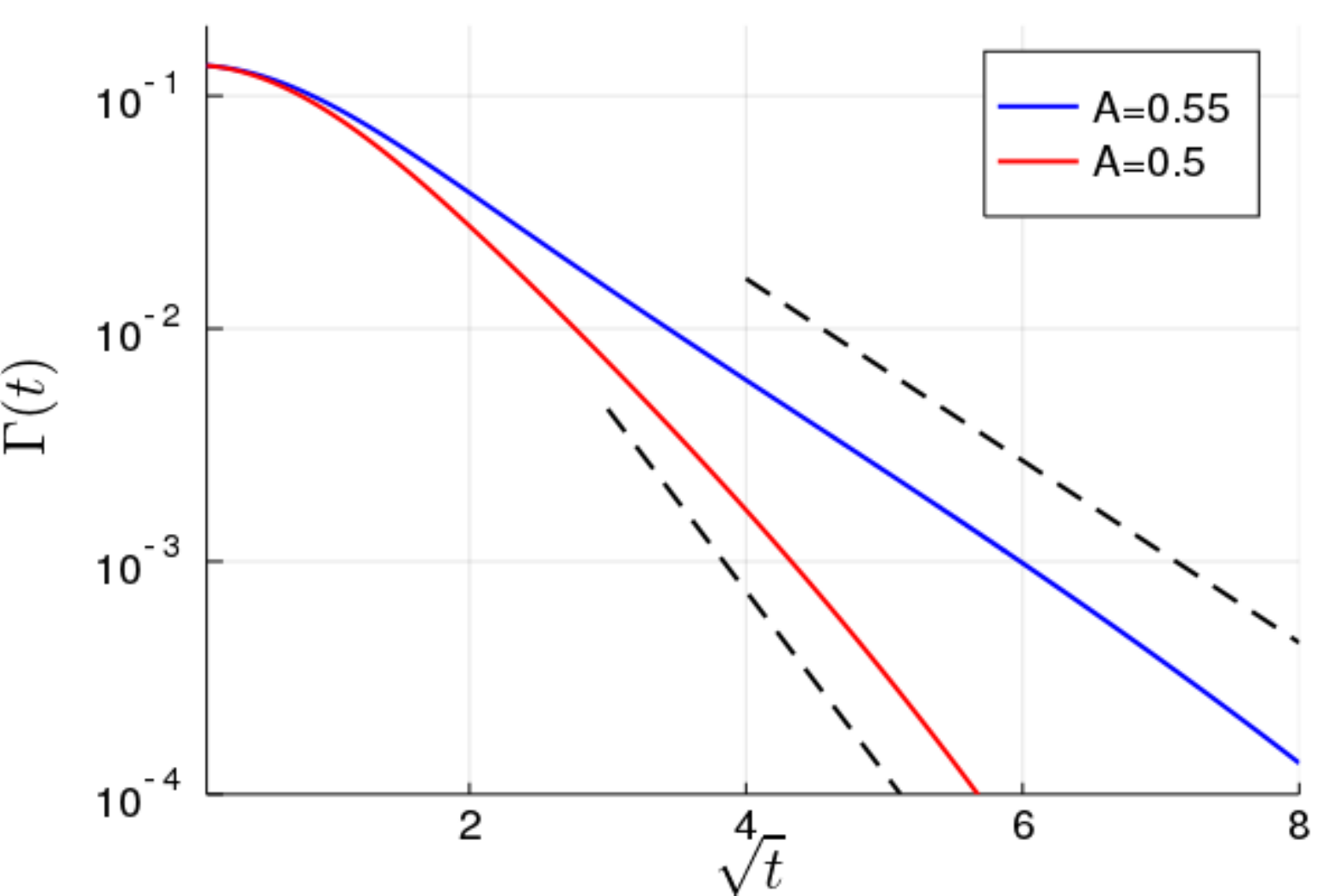}
\caption{\label{fig:sqrt_expo}Decay of $\Gamma(t)$ against $\sqrt{t}$, at $A=0.55$ and $A=0.5$ starting at the steady state with $\Gamma=0.134$ for $\mu=1$. Dashed lines show the predicted exponential decay with $\sqrt{t}$.}
\end{figure}

\begin{figure}
\includegraphics[scale=0.45]{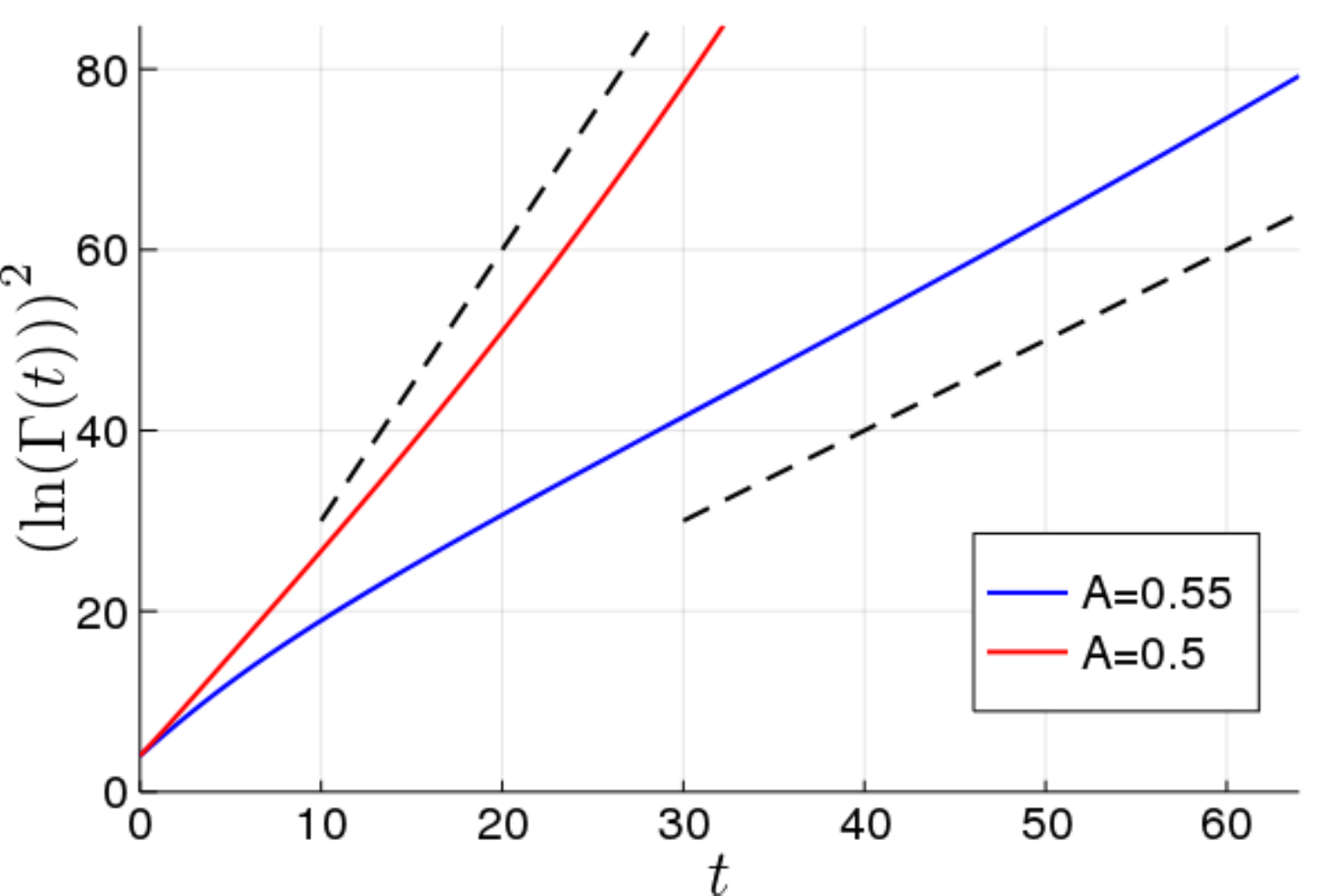}
\caption{\label{fig:log_square}Same data as in Figure \ref{fig:sqrt_expo}, with the same x and y ranges but now plotting ${(\ln(\Gamma(t)))}^2$ against $t$. This shows more clearly the linear growth (dashed lines) as predicted by equation (\ref{stretched}).}
\end{figure}

In the following we show the result of numerical simulations, where starting from our reference steady state we study the evolution at two values of $A$ below the transition, $A=0.55$ and $A=0.5$ (see Fig.~\ref{fig:aging_points}). We choose these higher values of $A$ for $\mu=1$ because for smaller values the yield rate becomes small very rapidly and the stretching regime is then difficult to resolve clearly. In the first plot (Fig.~\ref{fig:sqrt_expo}), we graph the yield rate in logarithmic scale against $\sqrt{t}$ and obtain a good agreement with the predicted straight line. In the second plot (Fig.~\ref{fig:log_square}) we show instead ${(\ln(\Gamma(t)))}^2$, which makes it easier to discern the asymptotic region with its linear increase in time.

 \subsection{\label{subsec:level3}Exponential decay for $\mu<1$}
 
\begin{figure}[th]
\includegraphics[scale=0.45]{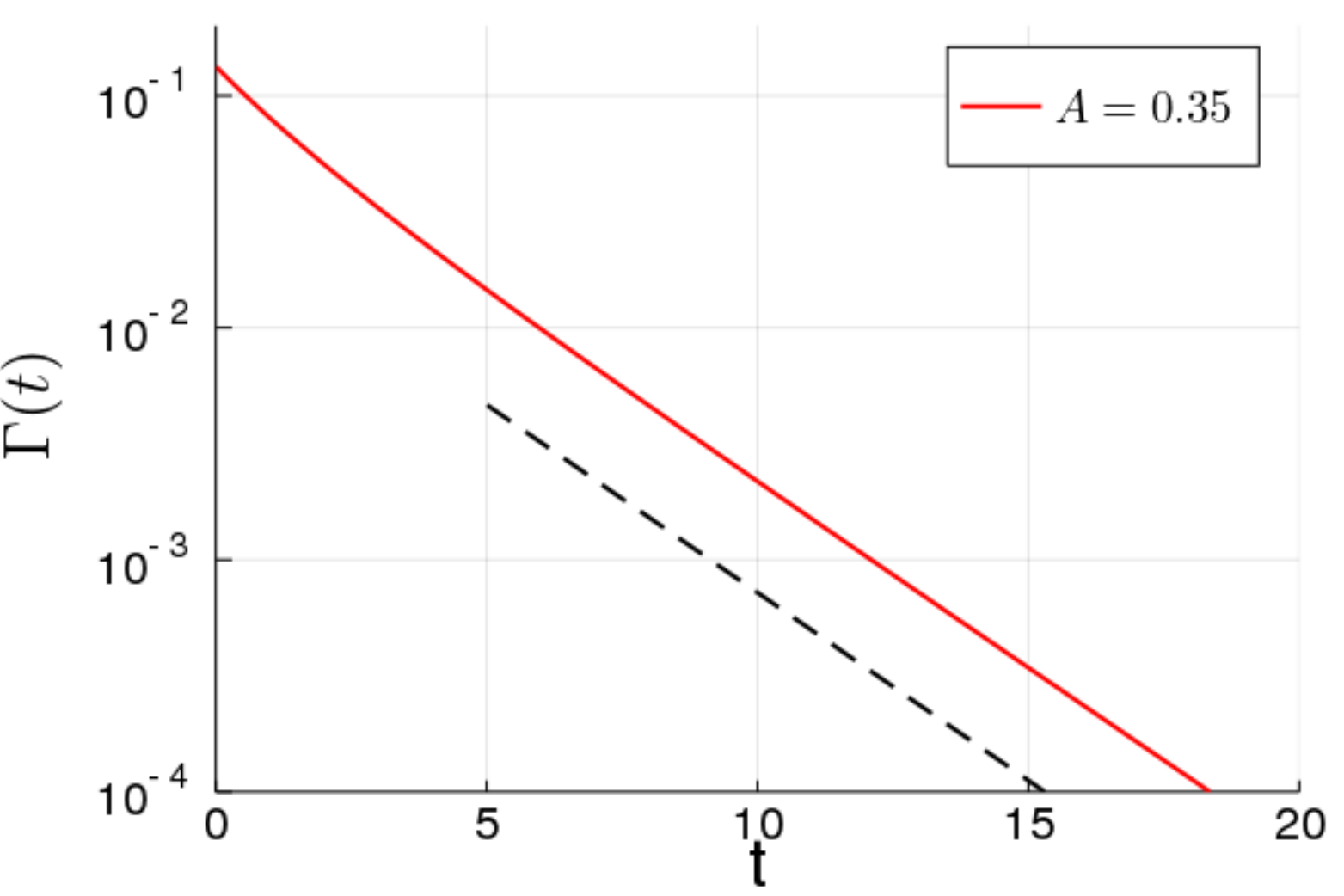}
\caption{\label{fig:gamma_mu_05}Yield rate decay at $A=0.35$, starting from a liquid-like steady state, for $\mu=0.5$. The dashed curve shows the exponential decay of $\Gamma(t)$.}
\end{figure}

In this regime we have not carried out a full boundary layer analysis. From a physical perspective, we expect it to be less relevant:
in contrast to $\mu=1$ (which is obtained directly from the elastoplastic decay of the propagator) and $1<\mu<2$ (for which there are coarse-graining arguments as in Refs.~\onlinecite{ferrero_criticality_2019,fernandez_aguirre_critical_2018}), there seems to be little evidence for this kind of noise distribution in real systems. We therefore show here only the results of a numerical evaluation. These are consistent with a purely exponential decay, as can be seen in Fig.~\ref{fig:gamma_mu_05} for $\mu=0.5$, where the system  evolves at $A=0.35$ starting from the reference steady state (see Fig.~\ref{fig:aging_points}).

\section{\label{sec:lattice}Aging in a lattice elastoplastic model}

In this final section we compare our mean field results to the aging behaviour found in a lattice elastoplastic model. This is important in order to test how well the mean field approximation, which discards spatial correlations, can approximate the full spatial dynamics.

The implementation of the elastoplastic model combines elastic loading and stochastic relaxation of a single element on each lattice site, with a spatially discretised fluid/continuum mechanical approach to enforcing Eshelby stress propagation after each stress drop, via the Stokes equation coupled to an additional elastic stress. On a timescale that scales with $\eta/\Go$, where $\eta$ is the viscosity and $\Go$ the elastic modulus of the local mesoscopic blocks, the Eshelby quadrupole in 2d is recovered, of course here in the form appropriate to a discrete square lattice with periodic boundary conditions.
To compare with our model we therefore take a value of the viscosity $\eta/G_0\ll 1$, as we had assumed -- in common with most elastoplastic models~\cite{nicolas_deformation_2018} -- that stress propagation after yielding takes place effectively instantaneously.

 \begin{figure}
 \hspace{0.5cm}
\includegraphics[scale=0.45]{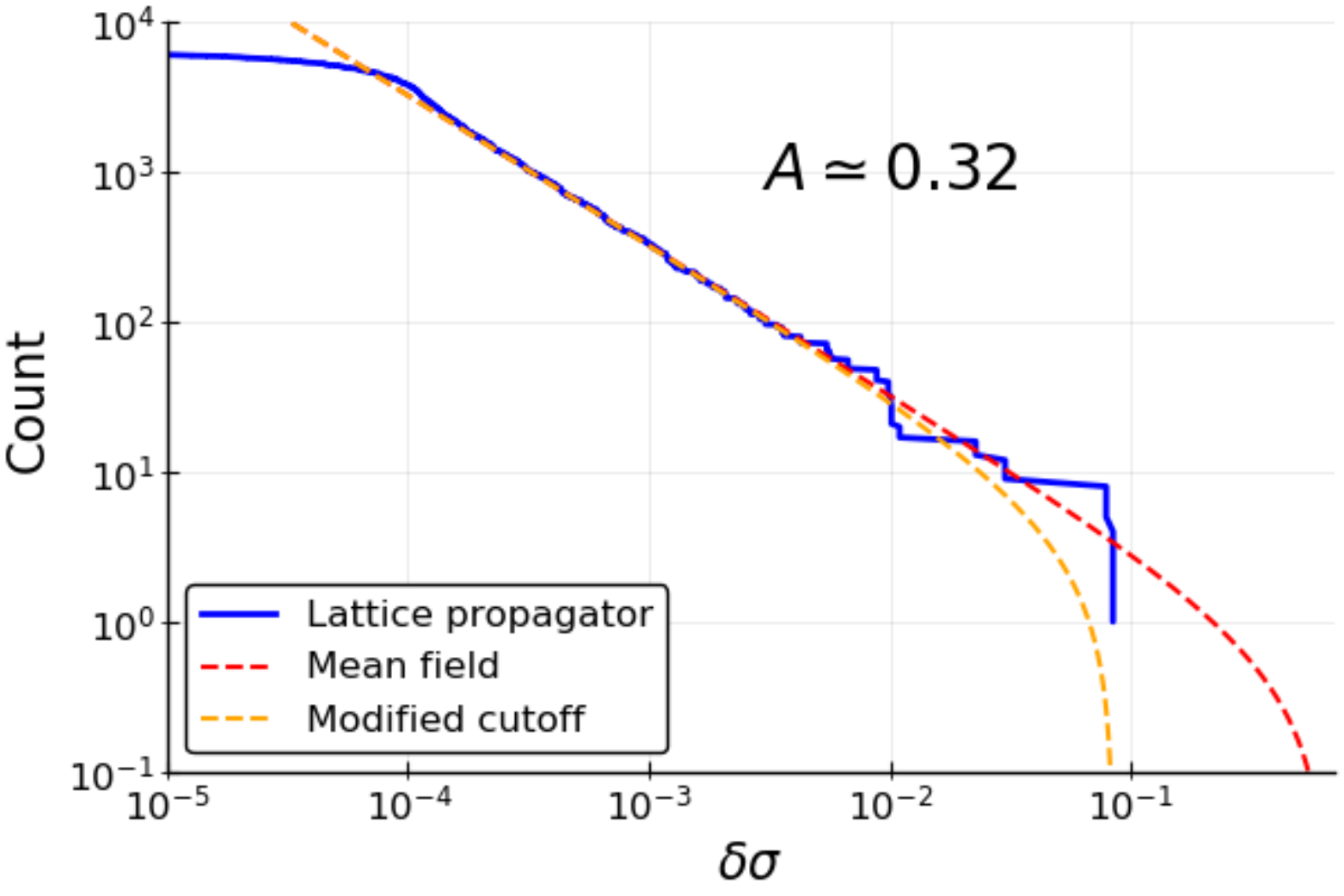}
\caption{\label{fig:counts}Cumulative count of the number of propagator elements larger than a given $\delta\sigma$, which drops to zero at the maximum stress kick. The lattice elements (blue) are compared to the model defined here (red) and to a power-law model with a modified cutoff (orange).}
\end{figure}

In order to connect our modelling approach with the lattice model, we first extract the corresponding value of $A$. To do so, we look at the list of $L^2$ stress propagator elements $\{\delta \sigma_i \}$ after a unit stress drop at the origin in a square lattice of size $L\times L$. Taking the positive elements, we sort them from largest to smallest and plot the ranking in the sorted list against the value of the element. This is shown in Figure~\ref{fig:counts}, and corresponds to a cumulative count of how many propagator elements are larger than a given $\delta\sigma$. For small $\delta \sigma$, where the power law behaviour (\ref{rho_wyart}) holds (see more below on the distribution at large $\delta \sigma$), this unnormalized cumulative distribution behaves as $P(x<\delta \sigma)\simeq A/\delta \sigma$. Therefore we can extract $A$ as the prefactor of the power law for small $\delta\sigma$ in Figure~\ref{fig:counts}, giving $A\approx 0.32$.

\begin{figure}
\includegraphics[scale=0.5]{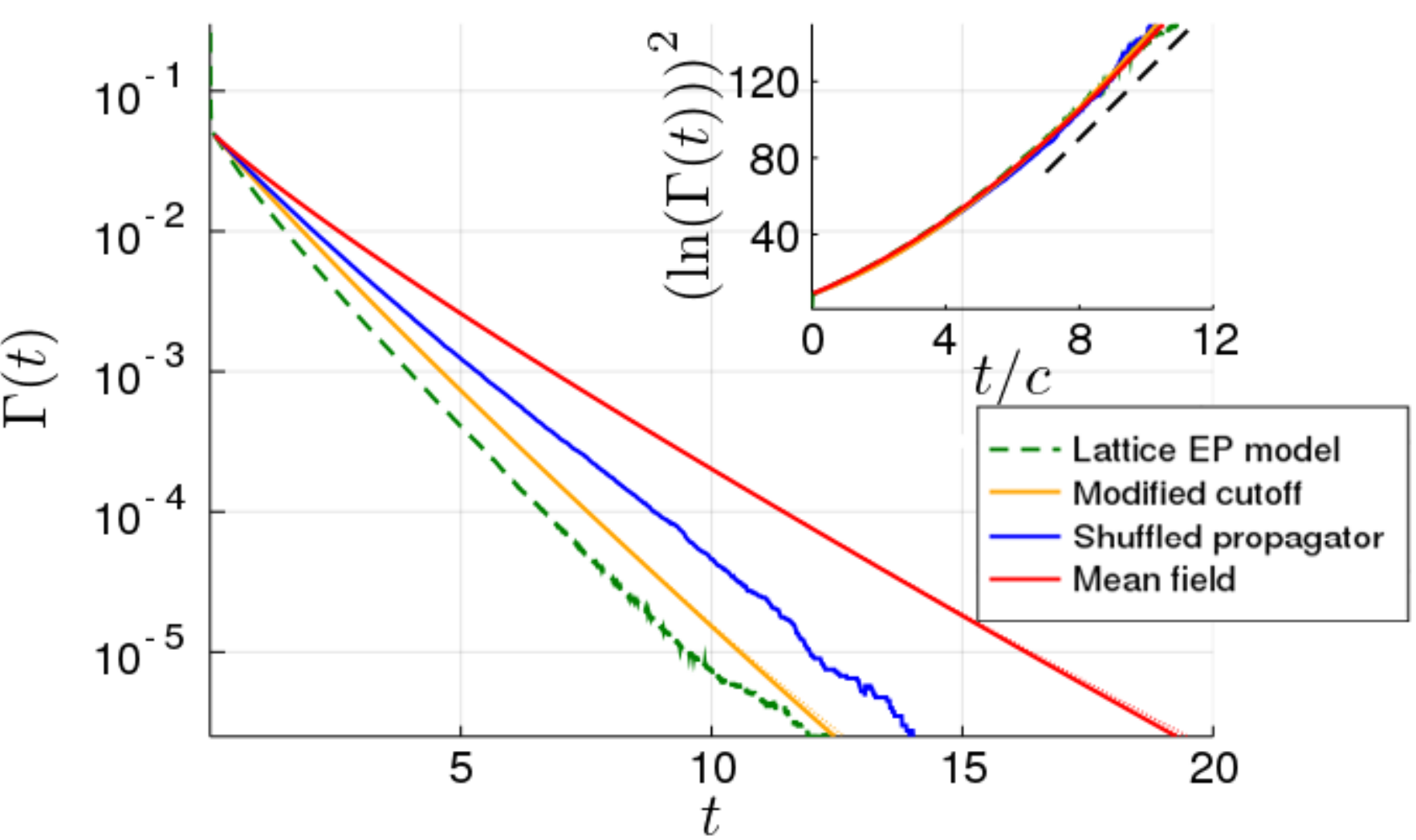}
\caption{\label{fig:without_legend}Yield rate decay for $A=0.32$ starting from the same initial distribution, for the full lattice model and the three mean field approaches described in the text. In the inset, the data are collapsed by rescaling time for each mean field prediction, and then fitted to a stretched exponential decay.}
\end{figure}

We can now run our mean field dynamics for $P(\sigma,t)$ with  the appropriate $A=0.32$, starting from the same initial stress distribution as in the lattice model (Figure~\ref{fig:without_legend}). The lattice data (green) is from a $4096 \times 4096$ system, where we can reliably measure yield rates down to around $\Gamma=5\times 10^{-6}$; at this point only $\mathcal{O}(10^2)$ unstable sites remain and finite size fluctuations become noticeable. In a first approach (red line), we compare this to the mean field model with $A=0.32$ with the definition of the upper cutoff in Section~\ref{sec:model_construction}. However, as can be seen already in Figure~\ref{fig:counts}, the large near-field lattice propagator elements are actually cut off at a smaller value on the lattice. In the lattice data fewer yield events are therefore triggered and the plastic activity $\Gamma(t)$ is lower, with a faster decay.

In a second approach, therefore, we consider the mean field model with a modified cutoff chosen as the largest propagator element on the lattice. This results in predictions (yellow line in Figure~\ref{fig:without_legend}) that are rather closer to the simulation results, but it obviously still neglects some details of the distribution of near-field propagator elements. This includes the fact that the lattice propagator list is not in fact entirely symmetric, e.g.\ because of the fact that the largest positive propagator elements occur for nearest neighbours while the largest negative ones (in the $\pm 45^\circ$ directions on the lattice) arise from next nearest neighbours. We therefore finally run (blue line in Figure~\ref{fig:without_legend}) the mean field theory using as our stress kick distribution the actual list of lattice propagator elements. This is implemented using a Gillespie algorithm that draws stress kicks randomly from this list~\footnote{For lattices of different sizes, the large $\delta \sigma$ details of the list of propagator elements are identical. In order to run Gillespie simulations with an arbitrary number of sites, we proceed by extrapolating the small $\delta \sigma$ power law in order to obtain the desired number of propagator elements.}. 
Such an approach still gives a somewhat slower decay of the yield rate than the full lattice run, in fact more so than the power law stress kick distribution with the modified cutoff (yellow line in Figure~\ref{fig:without_legend}).

More remarkable than the quantitative differences between the three implementations of the mean field description, however, is the very good qualitative agreement with the simulation data. In fact, with a simple rescaling of time for each mean field prediction we can achieve a near-perfect overlap with the simulation data, as shown in the inset of Figure~\ref{fig:without_legend}. The rescaling factors are modest ($c=1.7, 1.29, 1.12$, respectively for the three mean field approaches) but greater than unity, showing that the mean field ``reshuffling'' of stress propagation  tends to trigger more yield events across the system than in the underlying lattice model.

In the inset of Fig.~\ref{fig:without_legend} we plot the rescaled data in the same form as in Fig.~\ref{fig:log_square} to demonstrate that, importantly, the simulation data conform to the predicted stretched exponential asymptote. Surprisingly, though, even the pre-asymptotic yield rate decay is very well captured by the mean field theory.
These results support the idea that the propagator, which decays as $r^{-2}$ in $d=2$, is long range enough (see e.g.\ the discussion in Ref.~\onlinecite{fernandez_aguirre_critical_2018}) for our mean field predictions for the aging behaviour to apply almost quantitatively.

\section{\label{sec:outlook}Dicussion and outlook}

We have constructed in this paper a time-dependent mean-field model of amorphous solids, incorporating the power-law mechanical noise spectrum arising from localized plastic events~\cite{lemaitre_plastic_2007-1,lemaitre_plastic_2007,lin_mean-field_2016}. This model allows the exploration of general time-dependences, including aging and arbitrary rheological protocols. We have shown firstly the phase diagram of the model (in the absence of external shear), which separates the arrested (i.e.\ glassy) from the flowing (liquid) states. We then developed a boundary layer scaling approach, with the aim of studying the behaviour of the model at very low yield rates, as they appear e.g.\ during aging. As a first application, this allowed us to find the various scalings of the plastic rate in the (stationary) liquid regime just above the dynamical arrest transition. 

Our main findings concern the long time aging regime, the mean-field predictions for which we summarize here. We obtained three different regimes as the exponent $\mu=d/\beta$ characterizing the noise spectrum was varied. We recall that $\beta$ is a general decay exponent of the propagator $r^{-\beta}$, so that varying $\mu$ can be thought of as tuning the interaction range. We found that for $1<\mu<2$ the plastic activity $\Gamma(t)$ decays in a power-law fashion as $\sim t^{-\mu/(\mu-1)}$, reflecting the dominance of far-field events in determining the long-time dynamics. For $\mu<1$, on the other hand, near field events are dominant, and the relaxation becomes exponential. This is encoded in the boundary layer equation, where the crossing of the boundary is dominated by ``small" jumps in the first case and ``large" jumps in the second, giving rise to the \textit{homogeneous} and \textit{inhomogeneous} solutions respectively as described in Section~\ref{sec:boundary_layer}. In the marginal case $\mu=1$, where both near and far-field events are relevant, we found a stretched exponential decay of $\Gamma(t)$ that arises mathematically from the superposition of the two types of boundary layer solution. We compared this last case to simulations on a lattice elastoplastic model, and found a decay consistent with the predicted stretched exponential. In fact even the pre-asymptotic decay of the plastic activity is extremely well captured by the mean field predictions, up to a modest rescaling of time. This lends strong support to our mean field approach to the dynamics of amorphous solids.

We discuss next the meaning of the coupling constant $A$, which in the absence of shear (as considered in this work) is the only control parameter of the model once $\mu$ is given. Following the analysis of Section~\ref{sec:lattice}, where we extracted the value of $A=0.32$ from the lattice propagator elements, it may seem that the value of the coupling is fixed for a given system by its geometry. This is indeed the case if we consider our mean field model as derived from an elastoplastic model defined on a lattice. In Appendix~\ref{app:circular_geometry} we provide an alternative interpretation, based on Refs.~\onlinecite{lemaitre_plastic_2007-1,lemaitre_plastic_2007}. There a two-dimensional glassy system was studied under quasistatic shear, and a phenomenology was presented in terms of ``active" zones corresponding to shear transformation zones, which are distributed randomly throughout the system and can ``flip" leading to a plastic event. With this picture in mind, one sees (Appendix~\ref{app:circular_geometry}) that the coupling $A$ is in fact proportional to the area fraction (in $d=2$) of the system occupied by the active zones. Therefore the coupling $A$ may in fact be thought of as being related to structural details of the system, setting the effective coupling between mesoscopic elements in a way that is reminiscent of the effective mechanical temperature that appears in the shear-transformation-zone (STZ)\cite{falk_dynamics_1998} or soft glassy rheology (SGR)\cite{sollich_rheology_1997} theories.

Related to the question of the meaning of $A$ is the interpretation of the unsheared ($\dot{\gamma}=0$) steady state $\ps (\sigma)$, which we have referred to above as the liquid regime because its behaviour under shear would be Newtonian, without a yield stress. In the work cited above~\cite{lemaitre_plastic_2007}, dissipative plastic events were studied in the athermal quasi-static (AQS) regime; but it was speculated~\cite{lemaitre_plastic_2007} that such transformation zones may also be present in the absence of shear and may be related to dynamical heterogeneities. Other studies~\cite{maier_emergence_2017} show that Eshelby-like stress correlations are also present on the fluid side of the glass transition. This is consistent with the mean field picture of a liquid having a finite rate $\Gamma$ of local yield events. In any case, the fact that a liquid steady state $\ps (\sigma)$ was chosen as initial condition for the aging is not essential: other initial states as generated e.g.\ by oscillatory pre-shear could be considered and would not affect the long-time aging regime we have characterized here.

Ideally one would like to go beyond mesoscopic models and compare to particle-based numerical simulations or experiments. Regarding the former, an interesting model athermal system to consider would be the repulsive soft sphere model investigated in Ref.~\onlinecite{chacko_slow_2019}. When this athermal system undergoes a quench, it was shown to exhibit power law relaxation, with ``hot spots'' reminiscent of Eshelby events. To compare the results quantitatively to ours would require a method for extracting the rate of plastic rearrangements $\Gamma(t)$ from such simulation data. This method would, in particular, have to be able to separate events that occur together in avalanches, which is a significant challenge (see e.g. the methods employed in Ref.~\onlinecite{candelier_dynamical_2010}). 

For quantitative comparisons with simulations or experiments, further effects may need to be incorporated in a mean field model.
These may include the appearance of a growing length scale during the aging dynamics: in the present model we have considered for simplicity a system of identical rearranging sites, whose properties remain constant in time. Likewise, the yield threshold is considered uniform among the sites, while in reality the thresholds may be heterogeneous. This effect could be modelled by extracting the thresholds from a distribution, as previously studied within the HL model.~\cite{agoritsas_relevance_2015,agoritsas_non-trivial_2017} Another simplification of our model is that we consider scalar stresses (shear component only), while of course in reality stresses are tensorial. The inclusion of normal stresses may affect e.g.\ the aging dynamics, although the good agreement we found above with the lattice simulations -- which do account for these normal stresses -- suggests that such effects would only change the qualitative behaviour.
Furthermore, we have assumed throughout that stress propagation after a yield event is instantaneous. This could be included in a mean field model along the lines of the approach taken in Ref.~\onlinecite{bouchaud_spontaneous_2016} for the HL model, where once a site becomes unstable it remains so during an additional finite (restructuring) time scale before its stress is set back to zero. In the lattice elastoplastic model one could study a similar effect by investigating the dynamics at different values of the viscosity. Lastly, a possible extension of the model would be to include thermally activated events, which could be done following a recent approach~\cite{popovic_thermally_2020} inspired by depinning models. It would be interesting to see how this would affect the athermal aging dynamics described here (which in the presence of a finite activation rate would no longer be towards a frozen state, leading instead to a steady state at very long times), possibly leading to temperature-dependent exponents as was found in a Lennard-Jones glass~\cite{warren_quench_2013}.

A further interesting direction to explore will be the interplay between aging and rheology. One could extend the analysis here to investigate the aging of the linear shear response, studying the stress decay after a small step strain while the system relaxes from a state with high plastic activity. This was done for the HL model in Ref.~\onlinecite{sollich_aging_2017}, where the stress relaxation was shown to decay incompletely to a $t_{\rm w}$-dependent value, $t_{\rm w}$ being the waiting time between the initial system preparation and the time where the step strain is applied. In the present model we also expect an incomplete relaxation, but with a non-trivial time-dependence, which in the physical case $\mu=1$ will arise out of the stretched exponential decay of the activity. In contrast to the problems described above concerning the measurement of $\Gamma(t)$, here the stress and strain quantities are clearly defined and measurable so that one could compare with simulations in e.g.\ the model athermal system of Ref.~\onlinecite{chacko_slow_2019}, or experimental results 
for microgel particle suspensions~\cite{purnomo_glass_2008,purnomo_linear_2006,purnomo_rheological_2007}, where linear viscoelastic moduli have been measured in the aging regime. 

Moving beyond the linear regime, with the time-dependent model we have derived in this paper we can also study general rheological protocols such as creep response, where the system is held at constant stress below or just above the macroscopic yield stress. A mesoscopic elastoplastic approach to the creep problem was taken already in Ref.~\onlinecite{liu_creep_2018}, where the HL model was used to study the response of ``aged" configurations. Importantly, these initial configurations were set up by hand as Gaussian stress distributions, with their inverse width acting as a proxy for the system age. To go beyond this one would like a full model to capture both the aging dynamics leading to the initial condition, as well as the ensuing creep response; this should be possible with our approach. Besides the symmetric aging dynamics described here, one could consider also the case of pre-shear within the time-dependent model we have introduced. This preparation protocol is frequently used in experiments, e.g.\ on creep in carbopol microgels\cite{lidon_power-law_2017,agarwal_signatures_2019}. As these systems are typically regarded as athermal they would provide an interesting experimental system to compare to.


%
%

%


\appendix

\section{\label{app:model_construction}Derivation of time-dependent model}

\subsection{\label{app:model_construction_reduction}Reduction to stress distribution dynamics}

We show here the details of the derivation of the dynamics (\ref{pde}) for the local stress distribution from the full $N$-body master equation (\ref{N_body}). In this master equation the spatial structure of the stress propagation has already been removed and replaced by i.i.d.\ stress kicks.  Correlations do remain in the transition kernel (\ref{rate}) for finite $N$ but will disappear 
as $N\rightarrow{\infty}$. We thus assume directly the factorization
\begin{equation}
    P(\usigma)=\prod \limits_{i}P_i(\sigma_i)
\end{equation}
We are then interested in finding the dynamics of the stress distribution 
\begin{equation}
P(\sigma) = \frac{1}{N} \sum_i \langle \delta(\sigma-\sigma_i) \rangle = \frac{1}{N} \sum_i P_i(\sigma)     
\end{equation}
Each local $P_i(\sigma_i)$ can be 
obtained by marginalising out the remaining variables,
\begin{equation}
    P_i(\sigma_i)=\int \{\prod_{j\neq i} {\rm d}\sigma_j\} P(\usigma)
\end{equation}
This is trivial to do with the drift term of the master equation (\ref{N_body}). From the transition rates $K_l$, on the other hand, we will obtain two terms, which we call $I_1$ and $I_2$. These correspond respectively to the case when $l=i$ (so that the site that is yielding is the site for which we are finding the marginal distribution) and when $l \neq i$. We thus have
\begin{eqnarray}\label{marginal_pde}
    \partial_t P_i(\sigma_i)&=& \nonumber
    \int \{\prod \limits_{j \neq i} \mathrm{d}\sigma_j \}\partial_t P(\usigma)\\
    &=&{}-\dot{\gamma} \partial_{\sigma_i}P_i(\sigma_i)+I_1+I_2 
\end{eqnarray}
To simplify matters, we will at first carry out the calculations without the counterterm enforcing zero net stress change. We will justify this explicitly at the end, but intuitively one may already expect that this term gives sub-leading corrections for large $N$. Indeed the sum over random positive and negative increments will scale as $\sum_{k\neq l}\delta \sigma_k\sim \sqrt{N-1}\sqrt{\langle \delta \sigma^2 \rangle}$, so that we are considering in the end a term of order $\sqrt{\langle \delta \sigma^2 \rangle}/\sqrt{N-1}$, which is negligible compared to a typical stress kick $\delta \sigma$ for $N\gg1$. Leaving out the counterterm and exploiting that the stress kicks are i.i.d.\ random variables, we can generically simplify the delta term in the rates (\ref{rate}) to
\begin{equation}\label{simplified}
    \left\langle \prod \limits_{j \neq l}\delta\left(\sigma_j-\left(\sigma_j'+\delta \sigma_j\right)\right)\right\rangle=\prod \limits_{j\neq l}\rho (\sigma_j-\sigma_j')
\end{equation}

We begin now with the term $I_1$, corresponding to the case $l=i$. This is given by
\begin{equation}\label{I1}
    I_1=
    \int
    \{\prod \limits_{j \neq i} \mathrm{d}\sigma_j 
    \}\mathrm{d}\usigma'\left[K_i(\usigma|\usigma')P(\usigma')-K_i(\usigma'|\usigma)P(\usigma)\right]
\end{equation}
where the rate is simply
\begin{equation}
    K_i(\usigma|\usigma')=\theta(|\sigma_i'|-1)\delta(\sigma_i)\prod \limits_{j\neq i}\rho (\sigma_j-\sigma_j')
\end{equation}
Carrying out the marginalisation (\ref{I1}), one obtains that
\begin{equation}
    I_1=\left (\int\mathrm{d}\sigma_i' \theta(|\sigma_i'|-1)P_i(\sigma_i')\right)\delta(\sigma_i)-\theta(|\sigma_i|-1)P_i(\sigma_i)
\end{equation}
Therefore, when the yielding takes place at the site $i$ itself we obtain the yielding and the reinjection terms in the master equation, as one would expect.

The other term, $I_2$, which corresponds to  yield events at sites $l \neq i$, must then give rise to the propagator in the master equation. We need to compute
\begin{equation}\label{I2}
    I_2=\sum \limits_{l \neq i}
    \int
    \{\prod \limits_{j \neq i} \mathrm{d}\sigma_j 
    \}\mathrm{d}\usigma'\left[K_l(\usigma|\usigma')P(\usigma')-K_l(\usigma'|\usigma)P(\usigma)\right]
\end{equation}
In both terms the integrals over $\mathrm{d}\sigma_k \mathrm{d}\sigma_k'$ for $k \notin \{i, l\}$ just give factors of unity while the integration over $\mathrm{d}\sigma_l \mathrm{d}\sigma_l'$ results in a factor $\int\mathrm{d}\sigma_l \theta(|\sigma_l|-1)P_l(\sigma_l)$. This leaves
\begin{multline}
    I_2=\sum \limits_{l \neq i}\left (\int\mathrm{d}\sigma_l \theta(|\sigma_l|-1)P_l(\sigma_l)\right) \times\\
    \int \mathrm{d} \sigma_i'\rho(\sigma_i-\sigma_i')[P_i(\sigma_i')-P_i(\sigma_i)]
\end{multline}
We divide and multiply this expression by $N$ and add the $l=i$ term to the sum as it will only give a negligible $\mathcal{O}(1/N)$ correction:
\begin{multline}
    I_2=\frac{1}{N}\sum \limits_{l}
    \left (\int\mathrm{d}\sigma_l \theta(|\sigma_l|-1)P_l(\sigma_l)\right) \times\\
    \int \mathrm{d} \sigma_i'N\rho(\sigma_i-\sigma_i')[P_i(\sigma_i')-P_i(\sigma_i)]
\end{multline}
where
\begin{equation}
    N\rho(\sigma_i-\sigma_i')=N\frac{A}{N}|\sigma_i-\sigma_i'|^{-1-\mu}=A|\sigma_i-\sigma_i'|^{-1-\mu}
\end{equation}
This is the propagator in the master equation; the lower cutoff $\sim N^{-1/\mu}$ of $\rho(\delta\sigma)$ is immaterial here as it becomes negligible for $N\gg 1$.
Defining then the yield rate as the  average
\begin{equation}
    \Gamma=\frac{1}{N}\sum \limits_{l}\int\mathrm{d}\sigma_l \theta(|\sigma_l|-1)P_l(\sigma_l)
    = \int\mathrm{d}\sigma \ \theta(|\sigma|-1)P(\sigma)
\end{equation}
and inserting the results for $I_1$ and $I_2$, equation (\ref{marginal_pde}) becomes
\begin{multline}
     \partial_t P_i(\sigma_i)=-\dot{\gamma} \partial_{\sigma_i}P_i(\sigma_i)+\\ A \Gamma\int_{\sigma_i-\etau}^{\sigma_i+\etau}\frac{P_i(\sigma_i')-P_i(\sigma_i,t)}{|\sigma_i-\sigma_i'|^{\mu+1}}\mathrm{d}\sigma_i'\\
     -\theta (|\sigma_i|-1)P_i(\sigma_i)+\Gamma \delta (\sigma_i)
\end{multline}
Summing over $i$ and dividing by $N$ then gives the master equation (\ref{pde}) in the main text.

We finally have to show that the term enforcing a zero net stress change after the yield event is indeed negligible for $N \gg 1$. Instead of the simplified form (\ref{simplified}) we take into account now the full form of the rates (\ref{rate}) including this enforcing term, and look for the corresponding contribution to $I_1$ (\ref{I1}) and $I_2$ (\ref{I2}). Carrying out firstly the integrals over $\mathrm{d}\sigma_j \mathrm{d} \sigma_j'$ for $j\neq i$, the delta terms over $j\neq i$ all give factors of $1$ as was the case previously. We are then left only to deal with the term concerning site $i$. Introducing Fourier transforms, we can rewrite the delta function defining this term in the following manner:
\begin{multline}
     \delta\left(\sigma_i-\left(\sigma_i'+\delta \sigma_i-\frac{1}{N-1}\sum \limits_{k \neq l}\delta \sigma_k\right)\right)=\\
     \int \frac{\mathrm{d}\lambda}{2\pi} \prod \limits_{k\neq l,i}e^{i\lambda\delta \sigma_k/(N-1)}e^{i\lambda \left(\sigma_i-\sigma_i'-\delta \sigma_i(1-1/(N-1)) \right)}
\end{multline}
This term now needs to be averaged over the i.i.d.\ stress increment distributions. Doing so we find that 
\begin{multline}\label{Corrections}
    \left\langle  \delta\left(\sigma_i-\left(\sigma_i'+\delta \sigma_i-\frac{1}{N-1}\sum \limits_{k \neq l}\delta \sigma_k\right)\right)\right\rangle=\\
    \int \frac{\mathrm{d}\lambda}{2 \pi} {\hat{\rho}\left(\frac{\lambda}{N-1}\right)}^{N-2}e^{i\lambda (\sigma_i-\sigma_i')}\hat{\rho}\left(-\lambda\left(\frac{N-2}{N-1}\right)\right)
\end{multline}
in terms of the characteristic function
\begin{equation}
    \hat{\rho}(s)=\int \mathrm{d}(\delta \sigma) \rho (\delta \sigma) e^{i \delta \sigma s}
\end{equation}
Expanding this characteristic function for small argument, one can write
\begin{equation}
    \hat{\rho}(s)=1-\frac{1}{2}\langle {\delta \sigma}^2 \rangle s^2 +\mathcal{O}(s^4)
\end{equation}
where the variance $\langle {\delta \sigma}^2 \rangle$ of the microscopic stress kick distribution is given by
\begin{multline}
    \langle {\delta \sigma}^2 \rangle=2\int_{\etau N^{-1/\mu}}^{\etau} \mathrm{d}(\delta \sigma)\rho(\sigma){\delta \sigma}^2\\
    =\frac{2A}{N(2-\mu)}\etau^{2-\mu}+\mathcal{O}(N^{-2/\mu})\equiv \frac{B}{N}+\mathcal{O}(N^{-2/\mu})
\end{multline}
The first term involving $\hat\rho$ in (\ref{Corrections}) can now be shown to scale as
\begin{equation}
    {\hat{\rho}\left(\frac{\lambda}{N-1}\right)}^{N-2}\sim 1+\mathcal{O}\left(\frac{B\lambda^2}{N^2}\right)
\end{equation}
The final factor in (\ref{Corrections}) can be expanded in a similar fashion
\begin{equation}
    \hat{\rho}\left(-\lambda\left(1-\frac{1}{N-1}\right)\right)=\hat{\rho}(-\lambda)+\frac{\lambda}{N-1}\hat{\rho}'(-\lambda)+\mathcal{O}\left(\frac{1}{N^2}\right)
\end{equation}
For large $N$ the correction terms can be neglected and (\ref{Corrections}) becomes simply
\begin{equation}
    \int \frac{\mathrm{d}\lambda}{2\pi} \hat\rho(-\lambda)e^{i\lambda(\sigma_i-\sigma_i')} = \rho(\sigma_i-\sigma_i')
\end{equation}
as it would be if we had neglected the enforcing term from the beginning. The above analysis shows that the leading corrections to this result are $\mathcal{O}\left({{1}/{N}}\right)$.


\subsection{\label{app:model_construction_HL}The H\'{ebraud}-Lequeux limit}

We here show in detail how the HL diffusive model is obtained in the limit $\mu \rightarrow 2^{-}$ of equation (\ref{pde}) in the main text. This means specifically that the stress propagation term in the second line must reduce to the diffusive form $\alpha \Gamma(t)\partial^2_\sigma P$ in (\ref{hl_equation}). To demonstrate this, we apply to this term the Kramers-Moyal expansion of a master equation, which has the form
\begin{equation}
    \frac{\partial P(\sigma,t)}{\partial t}=\sum \limits_{n=1}^{\infty} \frac{(-1)^{n}}{n !}\frac{\partial^n}{\partial \sigma^n}\left(a_n(\sigma)P(\sigma,t)\right)
\end{equation}
The coefficients $a_n(\sigma)$, which in general may depend on $\sigma$, are the jump moments, i.e.\ the moments of the change in $\sigma$ weighted by the corresponding transition rate. In our case these are independent of the original stress and given by
\begin{equation}
    a_n=\int (\delta \sigma)^n \Gamma \rho(\delta \sigma) \mathrm{d}(\delta \sigma)
\end{equation}
(We have dropped the time-dependence of $\Gamma$ here, for simplicity of notation.) In other words, the $a_n$ are obtained as the moments of the L\'{evy} kernel in (\ref{pde}), which explicitly reads 
\begin{equation}
    \Gamma \rho(\delta \sigma)= \frac{A\Gamma}{|\delta\sigma|^{1+\mu}}
\end{equation}
Bearing in mind the upper cutoff $\etau$ on $|\delta\sigma|$, the coefficients $a_n$ then take the following form for even $n=2,4,6,\ldots$ (for odd $n$ they are zero as the kernel is an even function):
\begin{eqnarray}\label{coeffs}
    a_n=2 A \Gamma \int_0^{\etau}{\delta \sigma}^{n-1-\mu}\mathrm{d}(\delta \sigma)&=& \frac{2A \Gamma}{n-\mu}\left(\frac{2A}{\mu}\right)^{n/\mu-1}
\end{eqnarray}
We will fix the second jump moment to a constant by setting $a_2=2\alpha\Gamma$, where $\alpha$ is equal to the $\alpha_{\rm{eff}}$ introduced in the main text
\begin{equation}
    \alpha=\frac{A}{2-\mu}\left(\frac{2A}{\mu}\right)^{2/\mu-1}
\end{equation}
Inverting this relation, one finds that for the second jump moment to stay fixed the prefactor $A$ must vary with $\mu$ as
\begin{equation}
    A(\mu)=\frac{\mu}{2}\left(2\alpha\,\frac{2-\mu}{\mu}\right)^{\mu/2}
\end{equation}
Inserting this into equation (\ref{coeffs}) we find that the coefficients of the Kramers-Moyal expansion take the form
\begin{equation}
    a_n=\Gamma \frac{\mu}{n-\mu}\left(2\alpha\frac{2-\mu}{\mu}\right)^{n/2}
\end{equation}
We can now take the limit $\mu \rightarrow 2^{-}$ at fixed $\alpha$. One sees from the previous expression that in this limit
\begin{equation}
    a_n \simeq \frac{2\Gamma}{n-2}\alpha^{n/2}(2-\mu)^{n/2}\rightarrow 0\quad \forall n \geq 4
\end{equation}
We have therefore shown that all coefficients of the Kramers-Moyal expansion except the second one vanish for $\mu\to 2$. The remaining $n=2$ term then gives exactly the diffusive contribution $(1/2)\partial_\sigma^2(a_2 P)=\alpha \Gamma \partial_\sigma^2 P$ that is used in the HL model to represent stress propagation.

\section{\label{app:scaling_ss}Derivation of steady state scaling}

Considering again the steady state condition (\ref{rescaled_pde}), we view the stress distribution as a perturbed version of the critical distribution $P_c(\sigma)$ defined in Section~\ref{sec:phase_diagram}, which to connect with the boundary layer ansatz we write as  $Q_0(\sigma)$, so that $P(\sigma,\Gamma)=Q_0(\sigma)+\delta P (\sigma,\Gamma)$. Inserting this into (\ref{rescaled_pde}) and using the steady state condition for $Q_0(\sigma)$ at $A=A_c$, one finds for $\delta P(\sigma)$:
\begin{equation}\label{delta_P}
A \mathcal{L}\delta P(\sigma)-\frac{\theta(|\sigma|-1)}{\Gamma}\delta P(\sigma)-\tilde{A}\delta(\sigma)+A\So(\sigma)=0
\end{equation}
where we have defined
\begin{equation}
    \tilde{A}\equiv \frac{A-A_c}{A_c}
\end{equation}
and $\So(\sigma)$ is given by
\begin{equation}\label{S_0}
	\So(\sigma)=\theta(|\sigma|-1)\int_{-1}^{1}\frac{Q_0(\sigma')}{|\sigma-\sigma'|^{1+\mu}}\mathrm{d}\sigma'
\end{equation}

To find the critical scaling, we consider as in the HL model the normalization condition. Due to the linearity of equation (\ref{delta_P}), we can split $\delta P$ into a negative part $-\Delta^{\rm{orig}}(\sigma,\Gamma)$ from the (negative) source $-\tilde{A}\delta(\sigma)$ at the origin, and a positive part $\Delta^{\rm{ext}}(\sigma,\Gamma)$ from $A\So(\sigma)$. One can then write the normalization condition as a condition on the  integrals of these functions:
\begin{equation}
    \int \delta P(\sigma,\Gamma)\mathrm{d}\sigma\overset{!}{=}0=-\int \Delta^{\rm{orig}}\mathrm{d}\sigma+\int \Delta^{\rm{ext}}\mathrm{d}\sigma
\end{equation}

To obtain the first of these integrals we note that $\Delta^{\rm orig}(\sigma)$ solves exactly the standard steady state condition (\ref{rescaled_pde}), except that the source term is smaller by a factor $\tilde{A}$. Thus 
$\int \Delta^{\rm orig}(\sigma)\mathrm{d}\sigma$ is given by $\tilde{A}$ times the mean lifetime $\mfpt (A,\Gamma)$ of an effective particle diffusing from the origin. The lifetime $\mfpt (A,\Gamma)$ may be approximated by the value $\mfpt(A_c,\Gamma\to 0)=1$ up to higher order corrections in the small quantities $\Gamma$ and $\tilde{A}$. These can be neglected for the purpose of finding the leading scaling relations, thus simplifying the normalization condition to
\begin{equation}\label{Norms}
    \int \Delta^{\rm{orig}}(\sigma,\Gamma)\mathrm{d}\sigma\simeq\tilde{A}=\int \Delta^{\rm{ext}}(\sigma,\Gamma)\mathrm{d}\sigma
\end{equation}
We may now find the critical scaling by considering only the term $\Delta^{\rm{ext}}(\sigma,\Gamma)$, which follows equation (\ref{delta_P}) without the term at the origin. Coming back to the interpretation in terms of first passage times, one may think of the integral over $\Delta^{\rm{ext}}$ as a $\Delta \tau ^{\rm{ext}}$. This is the extra time that a particle lives before yielding when $\Gamma>0$, where yielding is no longer instantaneous but happens (in the rescaled equation) at a finite but large rate $\mathcal{O}(1/\Gamma) \gg 1$.

We can now follow the ansatz introduced in Sec.~\ref{sec:boundary_layer} (without the frozen-in term $Q_0(\sigma)$) to write $\Delta^{\rm{ext}}(\sigma,\Gamma)$ piecewise in the interior, boundary layer and exterior regions. Firstly, for the exterior region, where we write $\Delta^{\rm ext}=\Gamma T_1(\sigma)$, we find from (\ref{delta_P}) to leading order that $T_1(\sigma)=AS_0(\sigma)$ (the first term is smaller by an order of $\Gamma$). 

Proceeding to the boundary layer region, where $\Delta^{\rm{ext}}(\sigma,\Gamma)=\Gamma^{c}R_1(z)$ (without the frozen part), we need to incorporate the source $\So(\sigma)$, in its limiting form that applies within the boundary layer. To deduce this form we introduce the scaling variable $y=(1-\sigma')/(\sigma-1)$ and use that for $\sigma-1 \ll 1$ the integral (\ref{S_0}) will be dominated by the singular behaviour $Q_0(\sigma')\simeq \E (1-\sigma')^{\mu/2}$ near the boundary:
\begin{eqnarray}
\So(\sigma)&=&\E (\sigma-1)^{-\mu/2}\int_{0}^{2/(\sigma-1)}\!\!\frac{y^{\mu/2}}{(1+y)^{\mu/2}}\,\rm{d}y
\\
&\simeq& 
    \E \,\betaf\!\left(\frac{\mu}{2},1+\frac{\mu}{2}\right)(\sigma-1)^{-\mu/2}
    \\
    &=&\Gamma^{-1/2}\E\, \betaf\!\left(\frac{\mu}{2},1+\frac{\mu}{2}\right) z^{-\mu/2}
\end{eqnarray}
Comparing with (\ref{S(z)}) we see that this is precisely $\Gamma^{-1/2}S(z)$. On the other hand, because of the matching between $\Gamma^{c}R_1(z)$ and $T_1(\sigma)$ for $z \gg 1$ and $\sigma-1\ll 1$ we find that $c=1/2$, as in Sec.~\ref{sec:boundary_layer}. Writing the propagation in terms of rescaled variables as in (\ref{R_1(z)}), and together with the yielding term, we obtain finally that $R_1(z)$ as defined above for the boundary layer behaviour of $\Delta^{\rm ext}(\sigma)$ follows precisely the original boundary layer equation (\ref{BL_eq}). 

Considering the solution of the boundary layer equation on the interior as described in Sec.~\ref{sec:boundary_layer}, one can now deduce the scaling with $\Gamma$ of $\Delta ^{\rm{ext}}(\sigma)$ in the interior region. The integral $\int \Delta^{\rm ext}(\sigma,\Gamma)\mathrm{d}\sigma$ is dominated by this interior region, so that via equation (\ref{Norms}) we are eventually led to the critical scaling of the plastic activity (\ref{A_scaling}).

\section{\label{app:scaling}Scaling of $P(\sigma,\Gamma)$ }

In this appendix we give further details on the expansion for $\Gamma \ll 1$ of the distribution $P(\sigma,\Gamma)$. As outlined in the main text, the basis of the analysis is the boundary layer ansatz, whereby $P(\sigma,\Gamma)$ is expressed in a piecewise manner in the different regions (see also below).  This ansatz must then be inserted into the master equation in order to obtain equations for the scaling functions. Restricting ourselves to the 1st order in the expansion, these are $R_1(z)$, $T_1(\sigma)$ and $Q_1(\sigma)$, for the boundary layer (from now on BL), exterior and interior regions respectively. The frozen-in distribution $Q_0(\sigma)$ corresponds to the critical distribution $P_c(\sigma)$ in the steady state scaling analysis (Sec.~\ref{sec:scaling}), whereas in the aging it is fixed indirectly by the initial condition. We further restrict the analysis to the symmetric case. In the aging setting, parity is conserved by the time evolution, so the stress distribution will be symmetric if the same is true of the initial distribution, i.e.\ $P_0(-\sigma)=P_0(\sigma)$. We can then just focus on the bulk ( $-1+\epsilon\leq \sigma\leq 1-\epsilon$) and the right hand side BL ($1-\epsilon < \sigma < 1+\epsilon)$ and tail ($\sigma \geq 1+\epsilon$).

We will concentrate on the regime $1<\mu<2$, from which the marginal case $\mu=1$ may be obtained as a limiting case. The BL ansatz for $P(\sigma,\Gamma)$ is
\begin{widetext}
  \begin{equation}
    P(\sigma,\Gamma) =
    \left\{
    \begin{array}{lr}
      Q_0(\sigma)+\Gamma^{1/\mu}Q_1(\sigma),& {\rm{for}}\quad -1+\epsilon\leq \sigma\leq 1-\epsilon \\
      \Gamma^{1/2}R(z),& {\rm{for}} \quad1-\epsilon < \sigma < 1+\epsilon \\
      \Gamma T_1(\sigma),& {\rm{for}} \quad \sigma \geq 1+\epsilon
    \end{array}
    \right\}
\end{equation}
\end{widetext}
where the scaling function in the BL is written in terms of the re-scaled stress variable $z=\Gamma^{-1/\mu}(\sigma-1)$.

There are some comments to be made on the ansatz above. In particular we have already inserted the final values $a=1/\mu$, $b=1$ and $c=1/2$ of the exponents introduced in the main text:
\begin{itemize}
    \item The $b=1$ exponent for the external tail is clear from the analysis below of the external equation (in \ref{app:scaling_external}), where the external part of $P(\sigma,\Gamma)$ will be shown to be given by $Q_0$ convolved with the L\'{e}vy kernel, and thus proportional to the intensity $A\Gamma$ of this L\'{e}vy propagator.
    \item The exponent $c=1/2$ comes from matching the asymptotic power law of the BL function on the exterior, $\Rext(z)$ for $z \gg 1$, with the exterior tail $T_1(\sigma)$ for $\sigma-1 \ll 1$. In this regime $\Gamma T_1(\sigma)\sim \Gamma(\sigma-1)^{-\mu/2}$, which has to match with $\Gamma^c R(z)\sim \Gamma^c z^{-\mu/2}$ thus entailing $c=1/2$.
    \item As already pointed out in Section~\ref{sec:boundary_layer} of the main text, the exponent $a=1/\mu$ similarly stems from the match-up of the asymptotic power law of the BL function on the interior, $\Rint(z)$ for $|z| \gg1$, with the interior function $Q_1(\sigma)$ for $1-\sigma\ll 1$.
\end{itemize}

As already explained in the main text, the BL function $R(z)$ can be split further into a frozen contribution from $Q_0(\sigma)$ and a nontrivial piece we denote by $R_1(z)$. We recall equation (\ref{Rz_split}):  $R(z)=\E(-z)^{\mu/2}\theta(-z)+R_1(z)$.

We can now proceed by writing the full equation of motion, separately for the three different regions in $\sigma$.

\subsection{\label{app:scaling_external}External tail}
The equation of motion (\ref{pde}) in the (positive) external region for $T_1(\sigma)$, which for convenience we divide once by $\Gamma$, reads (from now on we do not write the upper cut-off on the stress changes and leave this implicit in the power law kernel):
\begin{widetext}
\begin{multline}\label{external}
    \frac{1}{\Gamma}\partial_t P(\sigma,\Gamma)=\frac{\dot{\Gamma}}{\Gamma}T_1(\sigma)=A \bigg ( \int_{-1+\epsilon}^{1-\epsilon} \frac{\Gamma^{{1}/{\mu}}Q_{1}(\sigma')+Q_0(\sigma')}{|\sigma-\sigma'|^{\mu+1}} \mathrm{d}\sigma' +\int_{1-\epsilon}^{1+\epsilon}\frac{\Gamma^{{1}/{2}}R_1(\Gamma^{-1/\mu}(\sigma'-1))}{|\sigma-\sigma'|^{\mu+1}}\mathrm{d}\sigma' \\+  \Gamma^{{1}/{2}}\int_{1-\epsilon}^{1} \frac{\E (-z')^{{\mu}/{2}}}{|\sigma-\sigma'|^{\mu+1}}\mathrm{d}\sigma'+    \int_{1+\epsilon}^{\infty}\frac{\Gamma(T_1(\sigma')-T_1(\sigma))}{|\sigma-\sigma'|^{\mu+1}}\mathrm{d}\sigma' -\Gamma T_1(\sigma)\int_{-\infty}^{1+\epsilon} \frac{1}{|\sigma-\sigma'|^{\mu+1}}\mathrm{d}\sigma'\bigg)
    -T_1(\sigma)
\end{multline}
\end{widetext}
The leading order terms in this equation are $O(1)$, and accordingly we have omitted incoming terms from the other side of the domain, i.e.\ $\sigma'<-1+\epsilon$, which vanish as $\Gamma\to 0$. The term from the time derivative $\dot{\Gamma}/\Gamma$ is also of lower order for any $\Gamma(t)$ decaying slower than an exponential (this is true both for the power-law decay for $1<\mu<2$ and the stretched exponential at $\mu=1$).

Taking into account the asymptotic power-law forms of the interior $\Rint(z)$ and exterior $\Rext(z)$ boundary layer function, one sees that the integral involving $R_1(z)$ is dominated by the upper and lower end. The dominance of the integral by its power-law asymptotes means that the shape of $R_1(z)$ inside the boundary layer is irrelevant here, thus we can simply extrapolate $T_1$ from the outside and $Q_{1}$ from the inside into  the boundary layer (i.e.\ consider $\epsilon \rightarrow 0$). Likewise, the contribution from the $\mu/2$ power-law term can be seen as just being the extension of the integral over the frozen-in distribution $Q_0(\sigma)$ all the way up to the boundary.

Collecting the leading order terms one therefore has
\begin{equation}\label{T}
    T_1(\sigma)=A\int_{-1}^{1}\frac{Q_0(\sigma')}{(\sigma-\sigma')^{\mu+1}}\mathrm{d}\sigma'
\end{equation}
So the external function is simply given by the convolution of the frozen-in distribution $Q_0(\sigma)$ with the L\'{e}vy kernel. Physically, this means that local stresses above the yield threshold typically arise when sub-threshold elements, with stresses following the frozen-in distribution $Q_0(\sigma)$, are perturbed by a single stress kick. We can evaluate the convolution explicitly in the limit where $\sigma-1 \ll 1$, which will have to match up with the BL function $R_1(z)$ for $z\gg 1$. Changing variable to $y=(1-\sigma')/(\sigma-1)$, and using that $Q_0(\sigma')\sim \E(1-\sigma')^{{\mu}/{2}}$, we can rewrite (\ref{T}) as
\begin{eqnarray}\label{T_approx}
    T_1(\sigma)&=&A \E(\sigma-1)^{-\mu/2}\int_{0}^{\frac{2}{\sigma-1}}\frac{y^{\mu/2}}{(1+y)^{\mu+1}}\mathrm{d}y\nonumber\\
    &\simeq& A \E \,\mathrm{B}\!\left(\frac{\mu}{2},1+\frac{\mu}{2}\right) (\sigma-1)^{-\mu/2}
\end{eqnarray}
which decays in the $\sim (\sigma-1)^{-\mu/2}$ fashion stated in the main text.

\subsection{\label{app:scaling_contributions}Contributions to the boundary layer}

The equation of motion (\ref{pde}) (again divided once by $\Gamma$) in the boundary layer region reads:
\begin{widetext}
\begin{multline}\label{BL}
    \frac{1}{\Gamma}\partial_t \tilde{P}(z,\Gamma)=\frac{1}{2}\Gamma^{-{1}/{2}}\frac{\dot{\Gamma}}{\Gamma}R_1(z)-\Gamma^{-{1}/{2}}\frac{\dot{\Gamma}}{\Gamma}R_1'(z)\frac{z}{\mu}=A\bigg (\int_{-1+\epsilon}^{1-\epsilon}\frac{\Gamma^{{1}/{\mu}}Q_{1}(\sigma')+Q_0(\sigma')}{|\Gamma^{1/\mu}z-(\sigma'-1)|^{\mu+1}} \mathrm{d}\sigma'\\+\int_{1+\epsilon}^{\infty}\frac{\Gamma T_1(\sigma')}{|\Gamma^{1/\mu}z-(\sigma'-1)|^{\mu+1}}\mathrm{d}\sigma'+\Gamma^{-{1}/{2}}\int_{-\epsilon \Gamma^{-{1/\mu}}}^{\epsilon \Gamma^{-{1}/{\mu}}}\frac{R_1(z')-R_1(z)}{|z-z'|^{\mu+1}}\mathrm{d}z' +\\\Gamma^{-1/2}\E\int_{-\epsilon \Gamma^{-{1/\mu}}}^{\epsilon \Gamma^{-{1/\mu}}} \frac{(-z')^{\mu/2}\theta(-z')-(-z)^{\mu/2}\theta(-z)}{|z-z'|^{\mu+1}}\mathrm{d}z'
     -\Gamma^{1/2}R_1(z)\int_{\mathrm{out}} \frac{1}{|\Gamma^{1/\mu}z-(\sigma'-1)|^{\mu+1}}\mathrm{d}\sigma'\bigg)-\theta(z)\Gamma^{-1/2}R_1(z)
\end{multline}
\end{widetext}
where we have already rewritten the integrals inside the BL in terms of the  scaling variable $z'=\Gamma^{-1/\mu}(\sigma'-1)$, and correspondingly expressed $\sigma$ in terms of $z$ as $\sigma=z \Gamma^{1/\mu}+1$. As in Appendix~\ref{app:scaling_external}, we again omit the lower order incoming terms from $\sigma'<-1+\epsilon$.
We have also introduced the notation $\int_{\rm{out}}$, to denote the integration over the region $(-\infty,1-\epsilon)\cup(1+\epsilon,+\infty)$.

The BL equation (\ref{BL_eq}) for $R_1(z)$ stated in the main text now follows from (\ref{BL}) using two arguments:
\begin{itemize}
    \item Firstly, the left hand side terms stemming from the time derivative are of lower order than the leading order terms $\mathcal{O}(\Gamma^{-1/2})$ in the equation. For the case $1<\mu<2$, where we will find that $\Gamma(t)$ decays in a power-law fashion as $\Gamma(t)\sim t^{-\mu/(\mu-1)}$, one has $\dot{\Gamma}/\Gamma=-\frac{\mu}{\mu-1}\Gamma^{(\mu-1)/\mu}$ and the time derivative terms are smaller by this factor than the leading $\Gamma^{-1/2}$ terms. In the case $\mu=1$, where we will find a stretched exponential, we also have that $\dot{\Gamma}\Gamma^{-3/2}\ll \Gamma^{-1/2}$ and so the time-derivative part is again of lower order and can be discarded in the final leading order equation for $R_1(z)$.
    \item Due to the way the boundary layer is defined, namely the fact that $\epsilon \Gamma^{-1/\mu}\gg 1$, it is simple to show that the terms propagating the bulk and exterior functions $Q_0$, $Q_1$ and $T_1$ into the boundary layer ($z=\mathcal{O}(1)$) are negligible. The bulk and exterior functions therefore only appear as matching conditions for the asymptotic power laws of $R_1(z)$. 
\end{itemize}

Overall, the left hand side of (\ref{BL}) can be discarded to leading order, as can the first and second terms on the right; in the third and fourth term the integration limits become $(-\infty,+\infty)$ and the fifth term is again subleading. Evaluating the $R_1$-independent fourth term explicitly then gives exactly (\ref{BL_eq}) in the main text.

\subsection{\label{app:scaling_interior}Equation in the interior}

The equation of motion (\ref{pde}) (divided once by $\Gamma$) in the interior region, where we write to leading order $P(\sigma,\Gamma)=Q_0(\sigma)+\Gamma^{1/\mu}Q_1(\sigma)$, reads 
\begin{widetext}
\begin{multline}\label{interior}
    \partial_t P=\frac{1}{\mu}\Gamma^{\frac{1}{\mu}-1}\frac{\dot{\Gamma}}{\Gamma}Q_{1}(\sigma)=A \bigg (\int_{1+\epsilon}^{\infty}\frac{\Gamma T_1(\sigma')}{|\sigma-\sigma'|^{\mu+1}}\mathrm{d}\sigma'+\int_{-\infty}^{-1-\epsilon}\frac{\Gamma T_1(-\sigma')}{|\sigma-\sigma'|^{\mu+1}}\mathrm{d}\sigma'
    +\int_{1-\epsilon}^{1+\epsilon}\frac{\Gamma^{\frac{1}{2}}R_1(\Gamma^{-\frac{1}{\mu}}(\sigma'-1))}{|\sigma-\sigma'|^{\mu+1}}\mathrm{d}\sigma'\\+\int_{-1-\epsilon}^{-1+\epsilon}\frac{\Gamma^{\frac{1}{2}}R_1(\Gamma^{-\frac{1}{\mu}}(-1-\sigma'))}{|\sigma-\sigma'|^{\mu+1}}\mathrm{d}\sigma'+
    +\Gamma^{\frac{1}{\mu}}\int_{-1+\epsilon}^{1-\epsilon}\frac{Q_{1}(\sigma')-Q_{1}(\sigma)}{|\sigma-\sigma'|^{\mu+1}}\mathrm{d}\sigma'+\int_{-1}^{1}\frac{Q_{0}(\sigma')-Q_{0}(\sigma)}{|\sigma-\sigma'|^{\mu+1}}\mathrm{d}\sigma'\\
    -(\Gamma^{\frac{1}{\mu}}Q_{1}(\sigma)+Q_0(\sigma))\int_{\rm out}\frac{1}{|\sigma-\sigma'|^{\mu+1}}\mathrm{d}\sigma'\bigg) +\delta(\sigma)
\end{multline}
\end{widetext}
where we again use the notation $\int_{\rm{out}}$, this time to denote the integration over the region $(-\infty,-1+\epsilon)\cup(1-\epsilon,+\infty)$.
It is clear to see in (\ref{interior}) that the leading order terms on the right hand side are $O(1)$, and are given by
\begin{multline}
    A\bigg(\int_{-1}^{1}\frac{Q_0(\sigma')-Q_0(\sigma)}{|\sigma-\sigma'|^{\mu+1}}\mathrm{d}\sigma'-Q_0(\sigma)\int_{\rm out}\frac{1}{|\sigma-\sigma'|^{\mu+1}}\mathrm{d}\sigma'\bigg)\\+\delta(\sigma)
\end{multline}
This term is nonzero for any $A<A_c$. To balance it and obtain a time-independent $Q_1$ we need the left hand side of (\ref{interior}) to be also of order $O(1)$, thus leading to the conclusion that $\dot{\Gamma}/\Gamma=O(\Gamma^{(\mu-1)/\mu}$). This implies that $\Gamma(t)$ decays in time in a power-law fashion as $\Gamma(t)=\prefQ t^{-b}$, with $b=\frac{\mu}{\mu-1}$, as stated in the main text. In particular one has $\dot{\Gamma}/\Gamma=-b t^{-1}=-b\Gamma^{1/b}\prefQ^{-1/b}$, where $\prefQ$ represents the prefactor of the power-law decay and can be extracted from the numerical solution of the dynamics. We therefore obtain an equation for $Q_1(\sigma)$
\begin{multline}\label{Q1}
    Q_1(\sigma)=-(\mu-1)\prefQ^{(\mu-1)/\mu}\Bigg( A\bigg(\int_{-1}^{1}\frac{Q_0(\sigma')-Q_0(\sigma)}{|\sigma-\sigma'|^{\mu+1}}\mathrm{d}\sigma'\\-Q_0(\sigma)\int_{\rm out}\frac{1}{|\sigma-\sigma'|^{\mu+1}}\mathrm{d}\sigma'\bigg)+\delta(\sigma)\Bigg)
\end{multline}
An example for this function is shown in Figure~\ref{fig:third_region_interior} in the main text.

\subsection{\label{app:scaling_mu_1}$\Rint(z)$ for $\mu=1$}

For the $\mu=1$ case, we will take the limit $\mu \rightarrow 1^{+}$ of equation (\ref{BL}), and perform an analysis similar to the one carried out in Ref.~\onlinecite{lin_microscopic_2018}. As mentioned above, the time derivative part on the left will be neglected, which will turn out be consistent at the end given the decay of $\Gamma(t)$.

We therefore have again the equation:
\begin{equation}\label{full_BL}
     A\int_{-\infty}^{\infty}\frac{R_1(z')-R_1(z)}{|z-z'|^{\mu+1}}\mathrm{d}z'+S(z)-\theta(z)R_1(z)=0
\end{equation}
and write the source $S(z)$ as
\begin{equation}
    S(z)=C_{S}z^{-\mu/2}, \quad C_{S}=\E \,\betaf\!\left (\frac{\mu}{2},1+\frac{\mu}{2}\right)
\end{equation}
For large positive $z$, the first (propagator) term is always subleading compared to $R_1$ itself so asymptotically $R_1$ has to exactly balance the source term:
\begin{equation}
    \Rext(z)=C_{\rm{ext}} z^{-\mu/2}
\end{equation}
with $C_{\rm{ext}}=C_{S}$. Next we consider the BL equation (\ref{full_BL}) for large negative $z$ ($z<0$, $|z|\gg 1$). We split the propagator term in (\ref{full_BL}) into three parts, which correspond respectively to jumps within the stable $z<0$ region, jumps out to the unstable $z>0$ region and incoming jumps from the unstable region (this term was negligible for large negative $z$ in the $\mu>1$ case):
\begin{multline}
    \int_{-\infty}^{\infty} \frac{R_1(z')-R_1(z)}{|z-z'|^{\mu+1}}\mathrm{d}z'=\int_{-\infty}^{0}\frac{\Rint(z')-\Rint(z)}{|z-z'|^{\mu+1}}\mathrm{d}z'\\
    -\Rint(z)\int_{0}^{\infty}\frac{1}{|z-z'|^{\mu+1}}\mathrm{d}z'+\int_{0}^{\infty}\frac{\Rext(z')}{|z-z'|^{\mu+1}}\mathrm{d}z'
\end{multline}
Using the asymptotic behaviour of $\Rext(z')$ for large positive $z'$, the last term with the jumps from the exterior part of the BL can be evaluated for large negative $z$ as
\begin{equation}\label{Source_fext}
    C_{\mathrm{ext}} \mathrm{B}(1-\mu/2,3\mu/2) |z|^{-3\mu/2}\equiv C_{+} |z|^{-3\mu/2}
\end{equation}
Given that we want to balance this term, and the convolutions with the power law kernel always reduce the power law exponent by  $\mu$, we write an ansatz for the interior BL behaviour as 
\begin{equation}
    \Rint(z)=C_{\mathrm{int}} |z|^{-\mu/2}
\end{equation}
With this, all terms in equation (\ref{full_BL}) scale as $|z|^{-3\mu/2}$ for large negative $z$. After appropriate rescalings, this equation then becomes
\begin{equation}
    \left(C_{\mathrm{int}} \int_{0}^{\infty}\frac{x^{-\mu/2}-1}{|1-x|^{\mu+1}}\mathrm{d}x-\frac{1}{\mu}C_{\mathrm{int}}+C_{+}\right)|z|^{-3\mu/2}=0
\end{equation}
From here we can find $C_{\mathrm{int}}$, which will be given by:
\begin{equation}\label{C_int_i}
    C_{\mathrm{int}}=\frac{C_{+}}{\frac{1}{\mu}-\int_{0}^{\infty}\frac{x^{-\mu/2}-1}{|1-x|^{\mu+1}}\mathrm{d}x}
\end{equation}
Given that this is the solution to the BL equation taking into account the source generated by jumps from the $z>0$ region (\ref{Source_fext}), we call it the \textit{inhomogeneous} solution and write it as
\begin{equation}
    \Rint{}^{\mathrm{i}}(z)=C_{\mathrm{int}}^{\mathrm{i}}|z|^{-\mu/2}
\end{equation}
where $C_{\mathrm{int}}^{\mathrm{i}}$ is given by equation (\ref{C_int_i}).

In Section~\ref{sec:boundary_layer} of the main text the exponent of the \textit{homogeneous} solution was worked out (the second exponent in Eq.~(\ref{two_behaviours})). The full solution (always in the regime of large negative $z$) may then be written as a superposition
\begin{equation}
    \Rint(z)=\Rint{}^{\mathrm{h}}(z)+\Rint{}^{\mathrm{i}}(z)=C_{\mathrm{int}}^{\mathrm{h}}|z|^{-(1-\mu/2)}+C_{\mathrm{int}}^{\mathrm{i}}|z|^{-\mu/2}
\end{equation}
For the purposes of taking the limit $\mu \rightarrow 1^{+}$ we introduce $\delta=\mu-1$, so that the limit becomes $\delta \rightarrow 0^{+}$. In this limit, the (negative) denominator of (\ref{C_int_i}) goes to zero linearly in $\delta$ (the exact prefactor from numerics is consistent with $\pi^2/2$; we absorb this into a rescaled version $\hat{C}_+$ of  $C_+$). We can therefore write the solution when $\delta \ll 1$ in the following form:
\begin{equation} \label{superposition}
    \Rint(z)=C_{\mathrm{int}}^{\mathrm{h}}(\delta)|z|^{-(1-\delta)/2}-\frac{\hat{C}_{+}}{\delta}|z|^{-(1+\delta)/2}
\end{equation}
In order for this to stay finite as $\delta \rightarrow 0$, the diverging last term has to be cancelled to leading order. This entails, from equation (\ref{superposition}), that  
\begin{equation}
    C_{\mathrm{int}}^{\mathrm{h}}=C_{\mathrm{BL}}+\frac{\hat{C}_{+}}{\delta}
\end{equation}
and so the prefactor of the homogeneous solution picks up a positive divergence in order to balance the negative divergence in the inhomogeneous one.

We now take the limit $\delta \rightarrow 0^{+}$ in equation (\ref{superposition}), and work out the form of the boundary layer function $R_1(z)$ in the limit $z\rightarrow -\infty$. Inserting the form of $C_{\mathrm{int}}^{\rm h}$ we have just derived, this becomes
\begin{equation}\label{fint}
    \Rint(z)=|z|^{-1/2}\left((C_{\mathrm{BL}}+\frac{\hat{C}_{+}}{\delta})|z|^{\delta/2}-\frac{\hat{C}_{+}}{\delta}|z|^{-\delta/2}\right)
\end{equation}
We now expand in the exponent:
\begin{equation}
    |z|^{\delta/2}=e^{\delta/2 \ln(|z|)}=1+\frac{\delta}{2}\ln(|z|)+\mathcal{O}(\delta^2)
\end{equation}
and plug this expansion back into equation (\ref{fint}) to obtain
\begin{multline}
     \lim_{\delta \rightarrow 0}\Rint(z)=|z|^{-1/2}\left(C_{\mathrm{BL}}+\hat{C}_{+}\ln(|z|)\right)\\\overset{|z| \gg 1}{\approx} \hat{C}_{+}|z|^{-1/2}\ln(|z|)
\end{multline}
We therefore see that the interior tail of the boundary layer function acquires a logarithmic correction at $\mu=1$.

\section{\label{app:circular_geometry}Derivation of $\rho(\delta \sigma)$ for a circular geometry}

\begin{figure}
 \hspace{0.5cm}
\includegraphics[scale=0.48]{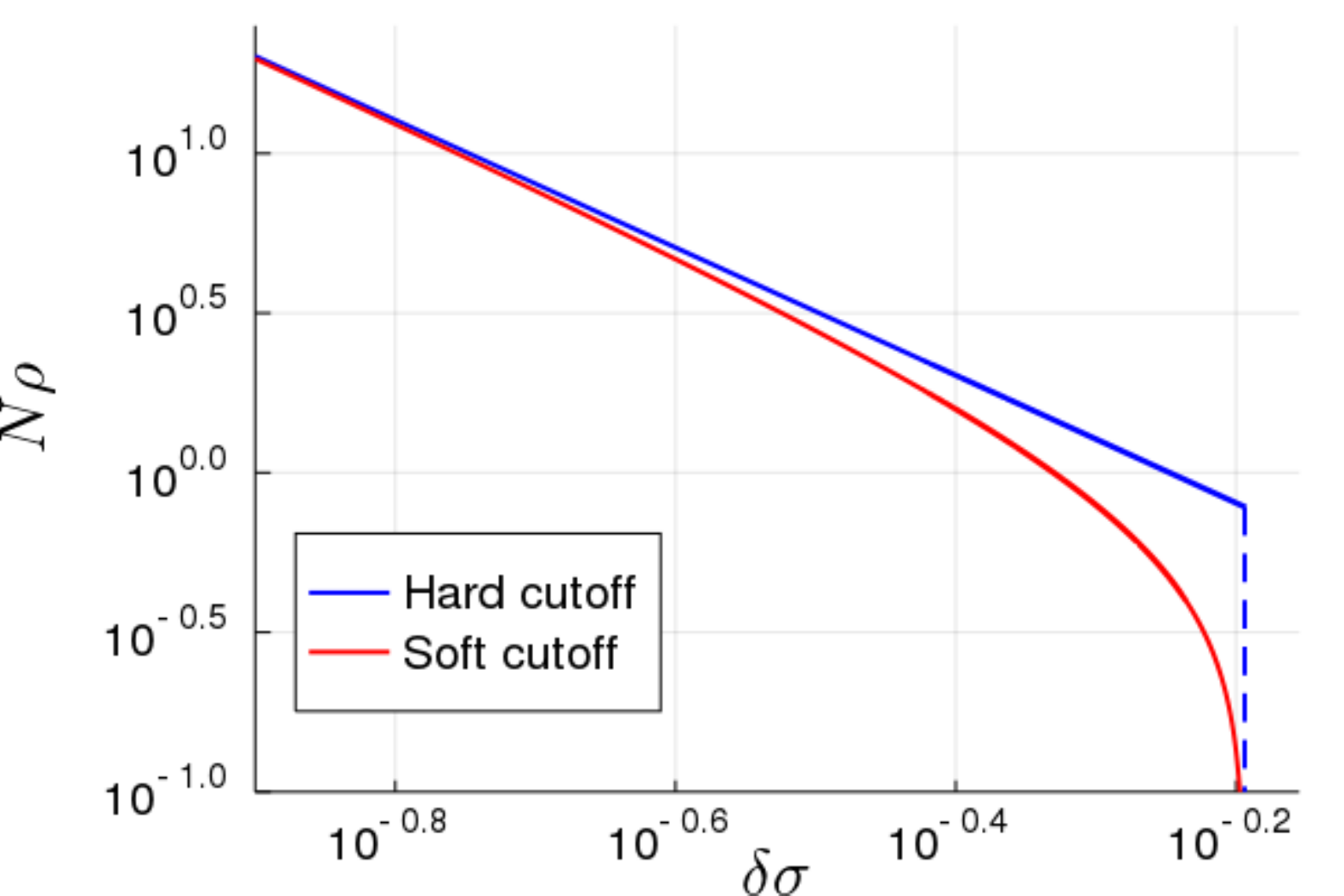}
\caption{\label{fig:hard_soft}Comparison of the hard cutoff expression (\ref{rho_wyart}) with the soft cutoff form (\ref{final_dist}). $A=0.32$ in the hard cutoff form; parameters in the soft cutoff version were chosen to have the same upper cutoff and power-law prefactor.}
\end{figure}

In this appendix we perform the explicit derivation of the mechanical noise spectrum $\rho(\delta \sigma)$ in the 2D case considering a circular geometry. In 2D, the stress field caused by a local plastic event at the origin reads in polar coordinates
\begin{equation}\label{propa}
    \delta \sigma(r,\theta)=\Go \aplas^2 \frac{\cos(4\theta)}{r^2}
\end{equation}
We then consider a circular geometry, so that the cluster of rearranging particles at the origin occupies a radius $\dplas$, while the rest of the mesoscopic elements lie uniformly within the ring around this, i.e.\ at a  distance $\dplas<r<\R$, $\R$ being the radius of the total system. The distribution over site \textit{positions} then reads
\begin{equation}\label{rtheta}
    \rho (r,\theta)=\frac{r}{\pi (\R^2-\dplas^2)} \mathrm{d}r\mathrm{d}\theta
\end{equation}
In order to obtain the distribution over stress increments $\rho (\delta \sigma)$, we need to perform the transformation $(r,\theta)\rightarrow \delta \sigma$ on the distribution (\ref{rtheta}), using the relation (\ref{propa}):
\begin{equation}
    \rho(\delta \sigma)=\frac{1}{\pi(\R^2-\dplas^2)}\int_{\dplas}^{\R}\!\!\mathrm{d}r\int_{0}^{2\pi}\!\!\mathrm{d}\theta \,r\,\delta\left(\delta \sigma-\Go\aplas^2\frac{\cos(4\theta)}{r^2}\right)
\end{equation}
We set $\theta'=4\theta$ so that the angular integration becomes $(1/4)\int_0^{8\pi}\mathrm{d}\theta'$; because of the periodicity of $\cos(\theta')$ we can then equivalently restrict the integration to $2\int_0^\pi \mathrm{d}\theta'$.
We now focus first on the positive half of the distribution ($\delta\sigma>0$), which corresponds to $\theta'<\pi/2$; the negative half may be obtained by symmetry. 
Performing then the variable change $x=\cos(\theta')$, we have for $\delta \sigma>0$
\begin{multline}\label{full_integral}
    \rho(\delta \sigma)=\\
    \frac{2}{\pi(\R^2-\dplas^2)}\int_{\dplas}^{\R}\mathrm{d}r\int_{0}^{1}\mathrm{d}x \frac{r}{\sqrt{1-x^2}}\delta \left(\delta \sigma-\Go \aplas^2 \frac{x}{r^2}\right)
\end{multline}
Using the properties of the delta function, we can then rewrite 
\begin{equation}
    \delta \left (\delta \sigma-\Go \aplas^2 \frac{x}{r^2}\right)=\frac{r^3}{2\Go \aplas^2 x}\delta \left (r-\sqrt{\frac{\Go \aplas^2 x}{\delta \sigma}}\right)    
\end{equation}
We can now perform the integral over $r$ in (\ref{full_integral}):
\begin{eqnarray}
    \int_{\dplas}^{\R}r^4\delta \left(r-\sqrt{\frac{\Go \aplas^2 x}{\delta \sigma}}\right)\mathrm{d}r&=&\left(\frac{\Go \aplas^2 x}{\delta \sigma}\right)^2
\end{eqnarray}
provided that     
\begin{equation}
\frac{\delta \sigma \dplas^2}{\Go \aplas^2}< x <\min\left(1,\frac{\delta \sigma \R^2}{\Go \aplas^2}\right)\ ;
\end{equation}
otherwise the integral vanishes. 
Therefore the full integral (\ref{full_integral}) becomes
\begin{equation}
    \rho^{+}(\delta \sigma)=\frac{\Go \aplas^2}{\pi (\R^2-\dplas^2)}\delta \sigma^{-2}\int_{\frac{\delta \sigma \dplas^2}{\Go\aplas^2}}^{\min\left(1,\frac{\delta \sigma \R^2}{\Go \aplas^2}\right)}\frac{x}{\sqrt{1-x^2}}\mathrm{d}x
\end{equation}
Performing the last integral and making use of the symmetry $\rho(-\delta \sigma)=\rho (\delta \sigma)$, the final expression for the stress kick distribution reads
\begin{multline}\label{final_dist}
    \rho (\delta \sigma)=\frac{\Go \aplas^2}{\pi (\R^2-\dplas^2)}\delta \sigma^{-2}\times\\
    \left(\sqrt{1-\left(\frac{\delta \sigma \dplas^2}{\Go \aplas^2}\right)^2}-\sqrt{1-\left(\min\left(1,\frac{|\delta \sigma| \R^2}{\Go \aplas^2}\right)\right)^2}\right)
\end{multline}
This distribution has two main features with respect to the pure power law distribution used in the main text. Firstly, it goes to zero continuously as $\delta \sigma$ approaches the highest possible stress kick in the system $\Go \aplas^2/\dplas^2$, instead of presenting a hard cutoff (see Figure~\ref{fig:hard_soft}). Secondly, the system size-dependent lower cutoff is also no longer sharp; instead $\rho(\delta\sigma)$ drops smoothly to a nonzero value as
$|\delta \sigma|$ decreases   below $\Go \aplas^2/\R^2$.
In the limit $R\to\infty$ and for $\delta\sigma$ below the upper cutoff 
one recovers the expected $\delta \sigma^{-2}$ power law decay. 

Finally, we discuss the result (\ref{final_dist}) in connection to the form of $\rho(\delta \sigma)$ used in the main text, defined by (\ref{rho_wyart}) and (\ref{etau_wyart}). To explore this, we replace the physical $r^{-2}$ propagator decay (\ref{propa}) by $\Go a^{\beta} \cos (4\theta)/r^{\beta}$, with a general decay exponent $\beta$. If we then carry out the same steps as described above, we obtain in the power-law region
\begin{equation}\label{wyart_mu}
    \rho(\delta \sigma)=\frac{c_{\mu}\mu (\Go \aplas^{\beta})^{\mu}}{\pi (\R^2-\dplas^2)}|\delta \sigma|^{-1-\mu}
\end{equation}
where we have defined $\mu=d/\beta=2/\beta$ as in the main text. The constant $c_{\mu}$ arises from the angular integration and is equal to
\begin{equation}
    c_{\mu}=\int_0^{1}\frac{x^{\mu}}{\sqrt{1-x^2}}\mathrm{d}x=\frac{\sqrt{\pi}\,\Gamma \left(\frac{\mu+1}{2}\right)}{2 \Gamma \left(\frac{\mu}{2}+1\right)}
\end{equation}
so that $c_1$=1, while in the limit case $\mu=2$ one has $c_2=\pi/4$.

We now need to relate the area of the system to the number of mesoscopic elements or zones $N$. To do so we introduce a dimensionless \textit{packing density}
\begin{equation}
    \pack=\frac{N\dplas^2}{\R^2}
\end{equation}
so that the distribution (\ref{wyart_mu}) may be written as
\begin{equation}
    \rho (\delta \sigma)=\frac{A}{N}{|\delta \sigma|}^{-\mu-1}\quad \text{with} \quad A= \frac{c_{\mu}\pack  \mu}{\pi} \left(\frac{\Go \aplas^{\beta}}{\dplas^{\beta}}\right)^{\mu}
    \label{rho_from_circle}
\end{equation}
The maximum stress change in the system $\etau$ can also be written in terms of the coupling $A$, as

\begin{equation}
    \etau=\frac{\Go \aplas^{\beta}}{\dplas^{\beta}}=\left (\frac{A\pi}{c_{\mu}\pack \mu}\right)^{1/\mu}
\end{equation}
We see from (\ref{rho_from_circle}) that the strength of the coupling constant $A$ depends crucially on the packing density $\pack$. On a lattice, the coupling $A$ will be fixed once and for all by the geometry of the underlying grid. However, if one takes the view in Ref.~\onlinecite{lemaitre_plastic_2007} of a variable density of ``active'' zones, which may be related to an effective mechanical temperature regulating the activity in the system, it becomes meaningful to think of a tunable coupling constant $A$ depending on the state of the system. 

\section{\label{app:numerical}Numerical methods}

\subsection{\label{app:numerical_pseudospectral}Pseudospectral method}

In integrating the master equation (\ref{pde}) numerically, the main challenge is the treatment of the stress propagation term. This L\'{e}vy propagator takes a simpler form in Fourier space:
\begin{equation}
    \int_{\sigma-\etau}^{\sigma+\etau}\frac{P(\sigma')-P(\sigma)}{|\sigma-\sigma'|^{\mu+1}}\mathrm{d}\sigma'\xrightarrow{\mathcal{F}}- |k|^{\mu}H_{\mu}(k\,\etau)\hat{P}(k)
\end{equation}
with 
\begin{equation}\label{Hmu}
    H_{\mu}(y)=\int_{0}^{y}\frac{1-\cos{t}}{t^{\mu+1}}\mathrm{d}t
\end{equation}
For a L\'{e}vy flight without cutoff, i.e.\ $\etau \rightarrow\infty$, $H_{\mu}(k\,\etau)$ converges to $I_{\mu}=-\cos{(\mu\pi/2)}\Gamma(-\mu)$ for $\mu \neq 1$, and $\frac{\pi}{2}$ for $\mu=1$. On the other hand, $\hat{P}(k)$ is just the Fourier transform of the stress distribution
\begin{equation}
    \hat{P}(k)=\int_{-\infty}^{\infty} e^{-i k \sigma}P(\sigma)\mathrm{d}\sigma
\end{equation}
To evolve the master equation (\ref{pde}) numerically we can therefore proceed in the following manner. Firstly we set up a stress grid $\{\sigma_i\}$ ($i=1,\ldots,M$) of $M$ points in a domain $(-l,l)$, along with the corresponding $M$ points $\{k_i\}$ in Fourier space. The domain width $2l$ has to be wide enough to avoid the effect of periodic images; for the aging simulations $l=4$ was used. To set up the discrete Fourier components of the propagator we evaluate the integral (\ref{Hmu}) numerically at each $k_i$, using an adaptive quadrature to account for the diverging power law in the integrand.

We can then employ a pseudospectral method, where we evolve alternately in stress and in Fourier space, using the FFT algorithm to switch between the two. Namely the L\'{e}vy propagator is applied in Fourier space, while all the other updates are realised in stress space. Finally we note that in the aging simulations, where the dynamics slows down at long times, we use an adaptive time step, which is fixed so that the maximum relative change in the system $\max \limits_{i}\Delta P(\sigma_i)/P(\sigma_i)$ stays within the range $(5\times 10^{-4},1\times10^{-3})$. In this way the timestep grows as the dynamics becomes progressively slower.

\subsection{\label{app:numerical_Stanley}Discrete matrix}

For calculating e.g.\ the mean first passage time that we employ to determine the phase diagram, we used an alternative numerical approach that allows us to implement absorbing boundary conditions. 

For this we follow Ref.~\onlinecite{buldyrev_average_2001} and write down a discretized transition matrix of the propagator 
\begin{equation}
    A\int_{-l}^{l}\frac{P(\sigma')-P(\sigma)}{|\sigma-\sigma'|^{1+\mu}}\mathrm{d}\sigma'
\end{equation}
We will start by considering the off-diagonal formally divergent integral; the second term, which corresponds to the diagonal, will be incorporated later by imposing probability conservation. 

We define again a stress grid $\{\sigma_i\}$ ($i=1,\ldots,M+1$) of $M+1$ points in a domain $(-l,l)$ (including now both boundaries), with a corresponding stress discretization $\Delta \sigma=2l/M$. For the off-diagonal term, it will be useful to represent the kernel as a discrete derivative:
\begin{eqnarray}
|\sigma-\sigma'|^{-\mu-1}&\simeq& K(\sigma,\sigma')\\&\equiv& \frac{1}{\mu\Delta\sigma}\left[|\sigma-\sigma'|^{-\mu}-|\sigma-\sigma'+\Delta\sigma|^{-\mu}\right]\nonumber
\end{eqnarray}
We can then write the integral in discretized form as
\begin{equation}
A\Delta\sigma \sum_j P(\sigma_j)K(\sigma_i,\sigma_j)
\end{equation}
To implement the cutoff, we set $K=0$ for $|\sigma_i-\sigma_j|>\etau$. The diagonal term is then
\begin{equation}
K(\sigma_i,\sigma_i)=\frac{2}{\mu\Delta\sigma}\left[-(\Delta\sigma)^{-\mu}+(\etau+\Delta\sigma)^{-\mu}\right]
\end{equation}
This is the same discretization as used in Ref.~\onlinecite{buldyrev_average_2001}, except for the modification in order to account for the presence of the upper cutoff in the propagator $\etau$. Absorbing boundary conditions are implemented by considering the discrete matrix of size $(M+1)\times (M+1)$, which effectively sets to $0$ all other elements outside so that whenever the stress at a site is ``kicked" out of the region $(-l,l)$ it is removed~\cite{zoia_fractional_2007}. 

In Section~\ref{sec:phase_diagram} we use $l=1$, so that we can implement the absorbing boundary conditions at $\sigma=\pm \sigma_c=\pm 1$. To obtain the points in the phase diagram (Figure~\ref{fig:phase_diagram}), for a fixed $\mu$ we compute $\mfpt(A)$ and bisect in $A$ until we find $\mfpt(A_c)=1$. $\mfpt(A)$ is computed in the way detailed in Ref.~\onlinecite{buldyrev_average_2001}, where the discretization above is introduced precisely to tackle the problem of the mean first passage time of a L\'{e}vy flight in a domain with absorbing boundaries. It is worth commenting on the discretization error. As we commented in Figure~\ref{fig:phase_diagram}, this error becomes larger as $\mu \rightarrow 2^{-}$. As worked out in Ref.~\onlinecite{buldyrev_average_2001}, in the limit $M\rightarrow\infty$ the error decays as $M^{\mu-2}$ for $1<\mu<2$, whereas for $\mu<1$ this crosses over to $M^{-1}$. From the scaling with $M^{\mu-2}$ one sees that the method breaks down when $\mu \rightarrow 2^{-}$. Intuitively, this may be expected, as one would be attempting to approximate what is effectively a (local) second derivative by a power law.

To obtain the data in Figure~\ref{fig:scaling} (Section~\ref{sec:scaling}), we solve for $\Delta^{\rm{ext}}(\sigma)$ numerically, making use again of the above discrete matrix. In this case, however, where we are considering a small but finite $\Gamma\ll 1$, we use $l=2>1$ in order to capture the external loss term in Eq.~(\ref{delta_P}). We implement as above absorbing boundary conditions at $l=\pm 2$, but this does not affect the result, as $\Gamma \ll 1$ and we have checked that the distribution decays to zero well within the $\sigma \in(-2,2)$ region. 

\subsection{\label{app:numerical_BL}Boundary layer equation}

We discuss in this section the numerical solution of the BL equation (\ref{BL_eq}) in the main text. This is required to obtain the full form of $R_1(z)$ for finite $z$, beyond the asymptotic power laws for large $|z|$ that we determine analytically in the main text.

The major difficulty in solving equation (\ref{BL_eq}) is that one a priori needs to consider an infinite domain $z \in (-\infty,\infty)$. To overcome this we proceed in the following manner. We cut the infinite domain down to a finite interval $(-l,l)$, by exploiting the knowledge we have on the tails of $R_1(z)$. That is, we split the incoming term of the propagator as 
\begin{equation}
    \int_{-\infty}^{\infty}\frac{R_1(z')}{|z-z'|^{\mu+1}}\mathrm{d}z'=\int_{-l}^{l}\frac{R_1(z')}{|z-z'|^{\mu+1}}\mathrm{d}z'+\mathrm{tail}_1(z)+\mathrm{tail}_2(z)
\end{equation}
where the two tails correspond respectively to the integrals over $\int_{-\infty}^{-l}$ and $\int_{l}^{\infty}$. In the two domains $z<-l$ and $z>l$, we assume the asymptotic forms of $R_1(z)$, which are power laws with prefactors proportional to $R_1(\pm l)$. Overall one can therefore include the two tail terms into the 1st and last column of the discrete transition matrix $K(\sigma_i,\sigma_j)$.

\begin{figure}[th]
 \hspace{0.5cm}
\includegraphics[scale=0.45]{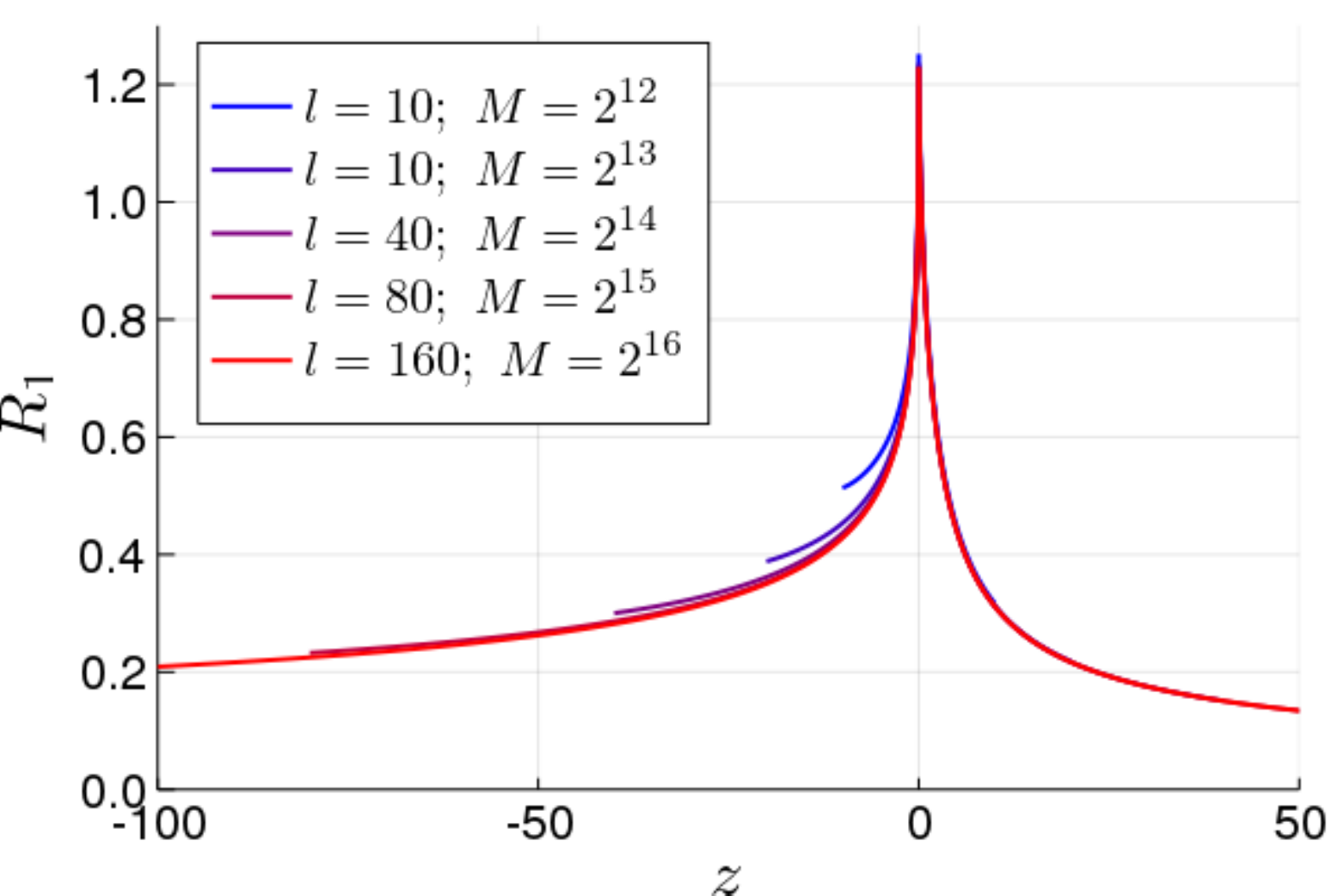}
\caption{\label{fig:BL_mu_1} Convergence of the solution of the BL equation (\ref{BL_eq}) with increasing size $l$ of the computational domain; parameter values are $\mu=1$, $A=0.58$ and source prefactor $\E=1$.}
\end{figure}

To check the convergence of the method, we can increase $l$ and $M$ (the number of grid points) while keeping the stress discretization $\Delta \sigma$ fixed. This is shown in Fig.~\ref{fig:BL_mu_1} for $\mu=1$, where we have assumed the asymptotic forms $|z|^{-1/2}\ln(|z|)$ on the left and $z^{-1/2}$ on the right. We can see that the method indeed converges as $l$ and $M$ are increased.

\section*{DATA AVAILABILITY}
The data that support the findings of this study are available from the corresponding author upon reasonable request.

\bibliography{ms}
\bibliographystyle{aipnum4-1}

\end{document}